\documentclass[%
reprint,
superscriptaddress,
pra,
amsmath,amssymb,
aps,
]{revtex4-1}

\usepackage{bm}
\usepackage[caption=false]{subfig}
\usepackage{braket}
\usepackage{amsfonts}
\usepackage[pdftex]{graphicx,color,hyperref}
\usepackage{booktabs,multirow}

\begin{document}
\title{Quantum remote sensing under the effect of dephasing}
\author{Hideaki Okane}
\email{h-okane@phys.kindai.ac.jp}
\affiliation{Nanoelectronics Research Institute, National Institute of Advanced Industrial Science and Technology (AIST)}
\affiliation{Department of Physics, Kindai University, Higashi-Osaka, 577-8502, Japan}
\author{Hideaki Hakoshima}
\affiliation{Device Technology Research Institute, National institute of Advanced Industrial Science and Technology (AIST), Central2, 1-1-1 Umezono, Tsukuba, Ibaraki 305-8568, JAPAN}
\author{Yuki Takeuchi}
\affiliation{NTT Communication Science Laboratories, NTT Corporation, 3-1 Morinosato-Wakamiya, Atsugi, Kanagawa 243-0198, Japan}
\author{Yuya Seki}
\affiliation{Device Technology Research Institute, National institute of Advanced Industrial Science and Technology (AIST), Central2, 1-1-1 Umezono, Tsukuba, Ibaraki 305-8568, JAPAN}
\author{Yuichiro Matsuzaki}
\affiliation{Device Technology Research Institute, National institute of Advanced Industrial Science and Technology (AIST), Central2, 1-1-1 Umezono, Tsukuba, Ibaraki 305-8568, JAPAN}
\date{\today}
\begin{abstract}
The quantum remote sensing (QRS) is a scheme to add security about the measurement results of a qubit-based sensor.
A client delegates a measurement task to a remote server that has a quantum sensor, and  eavesdropper (Eve) steals every classical information stored in the server side.
By using quantum properties, the QRS provides an asymmetricity about the information gain where the client gets more information about the sensing results than Eve.
However, quantum states are fragile against decoherence, and so it is not clear whether such a QRS is practically useful under the effect of realistic noise.
Here, we investigate the performance of the QRS with dephasing during the interaction with the target fields.
In the QRS, the client and server need to share a Bell pair, and an imperfection of the Bell pair leads to a state preparation error in a systematic way on the server side for the sensing.
We consider the effect of both dephasing and state preparation error.
The uncertainty of the client side decreases with the square root of the repetition number $M$ for small $M$,  which is the same scaling as the standard quantum metrology.
On the other hand, for large $M$, the state preparation error becomes as relevant as the dephasing, and the uncertainty decreases  logarithmically with $M$. 
We compare the information gain between the client and Eve.
This leads us to obtain the conditions for the asymmetric gain to be maintained even under the effect of dephasing.
\end{abstract}

\maketitle

\section{\label{sec:intro}Introduction}
Quantum properties such as a superposition and entanglement are considered as resource for information processing \cite{shor1999polynomial, PhysRevLett.79.325, PhysRevLett.103.150502, vandersypen2001experimental, BENNETT20147, bennett1992experimental, RevModPhys.74.145}.
Quantum computation provides a faster calculation than the classical one \cite{shor1999polynomial, PhysRevLett.79.325, PhysRevLett.103.150502, vandersypen2001experimental}.
Quantum cryptography guarantees a security during the transmission of information \cite{BENNETT20147, bennett1992experimental, RevModPhys.74.145}.
A hybrid architecture between these two schemes has been also discussed, which is called a blind quantum computation (BQC) \cite{broadbent2009universal, morimae2013blind, takeuchi2016blind, barz2012demonstration, greganti2016demonstration}.
The BQC provides a client with a way to delegate the quantum computation task  to a remote server.
Here, the client only has the primitive quantum device that cannot perform the full quantum computation, while the server has the quantum device to implement any quantum computation.
The important point of the BQC is to protect the privacy of the client's information such as input, output, and algorithm from the server.

A quantum sensor has been widely investigated as another application of quantum mechanics.
A superposition of a qubit acquires a relative phase affected by weak external fields, and we can efficiently obtain the information of the fields from the measurements on the qubit.
Such a qubit-based sensor is useful to detect magnetic field, electric field, or temperature \cite{degen2017quantum, budker2007optical, balasubramanian2008nanoscale, maze2008nanoscale, dolde2011electric, neumann2013high}.
Also, by using the entanglement between qubits, the sensitivity to the target field can be enhanced \cite{wineland1992spin, huelga1997improvement, matsuzaki2011magnetic, PhysRevLett.109.233601, Jones1166}.
Furthermore, the use of the quantum system at the nanoscale is expected to improve the spatial resolution of the target field \cite{maletinsky2012robust, schirhagl2014nitrogen}.

As a further development of quantum metrology, interdisciplinary approaches between quantum metrology and the other quantum technology have been discussed in Refs.~\cite{arrad2014increasing, kessler2014quantum, dur2014improved, herrera2015quantum, unden2016quantum, matsuzaki2017magnetic, higgins2007entanglement, waldherr2012high, nakayama2015quantum, matsuzaki2017projective, eldredge2018optimal, proctor2018multiparameter, giovannetti2001quantum, giovannetti2002quantum, PhysRevA.65.022309, PhysRevA.72.042338, komar2014quantum, PhysRevLett.98.120501, PhysRevA.99.022314, xie2018high}.
By using the quantum error correction which has been developed in the field of quantum computation \cite{lidar2013quantum},
the sensitivity to the target field in the presence of noise can be enhanced \cite{arrad2014increasing, kessler2014quantum, dur2014improved, herrera2015quantum, unden2016quantum, matsuzaki2017magnetic}.
Also, a quantum phase estimation 
algorithm \cite{kitaev1995quantum}, which can estimate a phase of an eigenvalue of a unitary operator $U$, is combined with the quantum sensing to improve the precision and the dynamic range of the sensor \cite{higgins2007entanglement, waldherr2012high}. 
Moreover, a quantum phase estimation 
algorithm provides a way to perform the projective measurements of energy without detailed knowledge of Hamiltonian \cite{nakayama2015quantum, matsuzaki2017projective}.
By constructing a network of the quantum metrology, the possibility to enhance the estimation precision at each location has been discussed in Refs.~\cite{eldredge2018optimal, proctor2018multiparameter}.
Beside these interdisciplinary approaches, when one sends the data obtained by the quantum metrology to the remote site, quantum cryptography is used for secure communication as in Refs.~\cite{giovannetti2001quantum, giovannetti2002quantum, PhysRevA.65.022309, PhysRevA.72.042338, PhysRevLett.98.120501, komar2014quantum, PhysRevA.99.022314, xie2018high}.

Recently, quantum remote sensing (QRS)~\cite{takeuchi2019quantum, yin2019experimental} has been proposed and demonstrated as a new interdisciplinary approach in which the concept of  BQC is applied to quantum metrology.
Similarly to the BQC, the QRS enables the client to delegate a task of the quantum metrology safely.
Here, the client has the device that can only measure a single-qubit quantum state, while the remote server has a sensitive quantum sensor to measure a target field.
Even if Eve attacks the server's device to steal every classical information recorded at the server, Eve should not obtain the information about the target fields.
For this purpose, a Bell state shared between the client and the server plays an important role for the protection of the information on the measured target field. 
Also, this protocol of the QRS is experimentally demonstrated with an optical setup in Ref.~\cite{yin2019experimental}.
However, an experimental demonstration of the QRS with solid-state systems has not been implemented yet.

In the actual experiment, the channel noise for the Bell state shared between the client and the server is inevitable, and the effect of the noise should be considered.
A random-sampling test is one of the ways to guarantee the quality of the Bell pair that was potentially damaged by such a noise channel.
The random-sampling test ensures the lower bound for the fidelity between the actual state and the Bell state generated in the experiment, as shown in Ref.~\cite{takeuchi2019quantum}.
This channel noise leads to a systematic error for the uncertainty of the qubit frequency $\delta\omega_{\mathrm{C}}$ for the client side
due to a deviation of the initial state in the quantum metrology of the server side.
Moreover, by assuming that Eve can know information on the error of the state preparation due to the channel noise, the ratio between the uncertainties of the client and Eve is evaluated in Ref.~\cite{takeuchi2019quantum}.

In this paper, we consider the effect of dephasing in the QRS.
Dephasing is one of the typical noise in the solid-state systems, and there are many previous researches about how dephasing affects the quantum sensing \cite{degen2017quantum, huelga1997improvement, matsuzaki2011magnetic, PhysRevLett.109.233601, Jones1166}.
On the other hand, in the QRS, we should consider not only the dephasing but also the systematic error caused by imperfect initial state preparation.
Due to the systematic error, the uncertainty $\delta\omega_{\mathrm{C}}$ does not decrease in proportion to $1/\sqrt{M}$ where $M$ denotes the number of the repetitions.
Additionally, the effect of dephasing degrades the coherence of the qubit, and so it is important to optimize the interaction time between the qubit and target magnetic fields, which is determined by a tradeoff relationship between other parameters.
In particular, due to the existence of the systematic error of the state preparation, such an optimization becomes highly non-trivial.
In the QRS protocol under the effect of dephasing, 
we have found that the optimized interaction time $t_{\mathrm{opt}}$ to minimize the uncertainty $\delta\omega_{\mathrm{C}}$ depends on the repetition number $M$ and the error rate of the state preparation $\epsilon$.
We investigate how the increase of the repetition number $M$ changes the behavior of the uncertainty $\delta\omega_{\mathrm{C}}$ with the optimized interaction time $t_{\mathrm{opt}}$.
Moreover, we calculate the uncertainty of the qubit frequency $\delta\omega_{\mathrm{E}}$ for the Eve, and compare this with  $\delta\omega_{\mathrm{C}}$.
Our results show the conditions for the asymmetric gain $\delta\omega_{\mathrm{C}}/\delta\omega_{\mathrm{E}}<1$ to be maintained even under the effect of dephasing.

The rest of this paper is organized as follows:
the section \ref{sec:rqsreview} reviews the QRS.
The section \ref{sec:model} introduces models of the dephasing and state preparation error.
The section \ref{sec:uncer} analytically calculates the uncertainties of the estimation for the client and Eve.
The section \ref{sec:opt} optimizes the interaction time to minimize the client's uncertainty, and shows the numerical results of the uncertainty for the client and the ratio between the uncertainties for the client and Eve.
The last section \ref{sec:Concl}  summarizes our results.

\section{\label{sec:rqsreview} Quantum remote sensing}
\begin{figure}
  \centering
  \includegraphics[width=8.5cm]{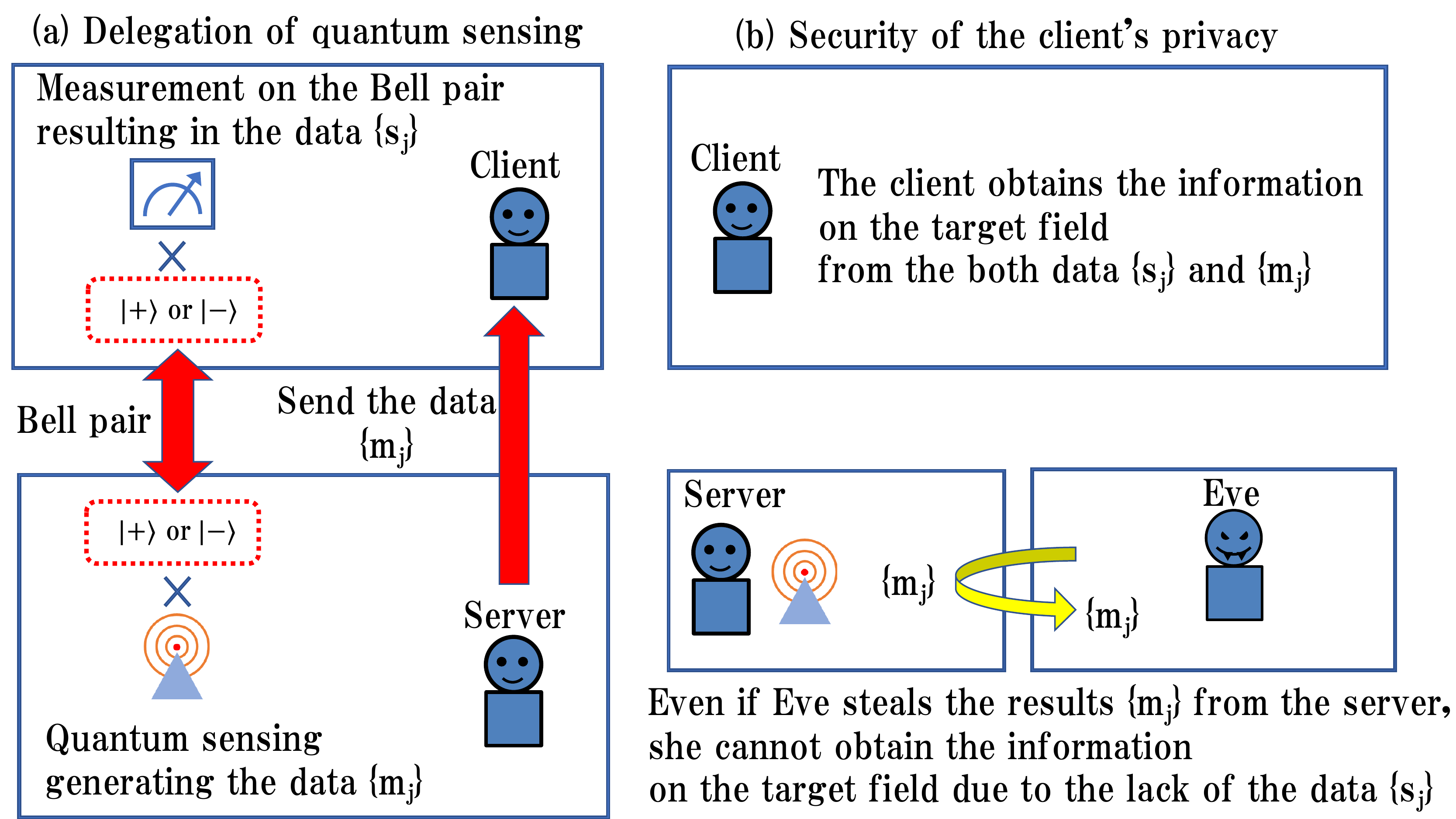}
  \caption{A schematic diagram to explain the outline of the QRS.
   We consider the case that the client delegates a task of measuring magnetic fields to the remote server with a quantum sensor.
   (a) First, after the Bell pair is shared between the client and the server, the client measures the Bell pair and obtains the data $s_j\in\{0,~1\}$ to identify the initial state of the quantum sensing ($\ket{+}~\text{or}~\ket{-}$) in the server side.
   The server performs the quantum sensing with the initial state described above, and sends the obtained data $m_j\in\{0,~1\}$ to the client.
   (b) By using the both data $\{s_j\}$ and $\{m_j\}$, the client can estimate the amplitude of target magnetic fields.
   On the other hand, even if Eve attacks the server side and steals the data $\{m_j\}$, she cannot estimate the target field due to the lack of the data $\{s_j\}$, which contains the information on the initial states prepared for performing the quantum sensing.
  }\label{flow}
\end{figure}

In this section, we explain the basic idea of the QRS as shown
in Fig.~\ref{flow}.
For simplicity, let us assume that a perfect Bell pair is available between the client and server (although we will relax this condition later). 
The flow of the QRS is as follows:
\begin{enumerate}
\item The Bell state $\ket{\Phi^+}\equiv(\ket{00}+\ket{11})/\sqrt{2}$ is shared between the client and the server.
\item The client measures his/her part of the Bell state by $\sigma_x$ base to prepare the initial state $\ket{+}$ or $\ket{-}$ for the server side where $\ket{\pm}\equiv(\ket{0}\pm\ket{1})/\sqrt{2}$.
\item The server performs the standard Ramsey type quantum metrology (as shown in Appendix \ref{app:a}) with the initial state $\ket{+}$ or $\ket{-}$ to measure the target field, and sends the results to the client.
\item Repeat the steps 1--3, $M$ times.
\end{enumerate}
Due to the $\sigma_x$ measurement of the Bell state by the client in the step 2, the information on the initial state for the quantum sensing is not known to the server.
On the other hand, both of the $\sigma_x$ measurement results at the step 2 and quantum sensing results at the step 3 are available for the client, and so the client can obtain the information on the target field.
On the other hand, even if Eve attacks the server side and steals the quantum sensing results, she cannot estimate the target field due to the lack of the information on the initial states.
In this sense, the QRS protocol certainly guarantees the privacy of the client under the condition that the Bell pair is prepared perfectly.


\section{\label{sec:model} noise model}
\begin{figure}
  \centering
  \includegraphics[width=8.5cm]{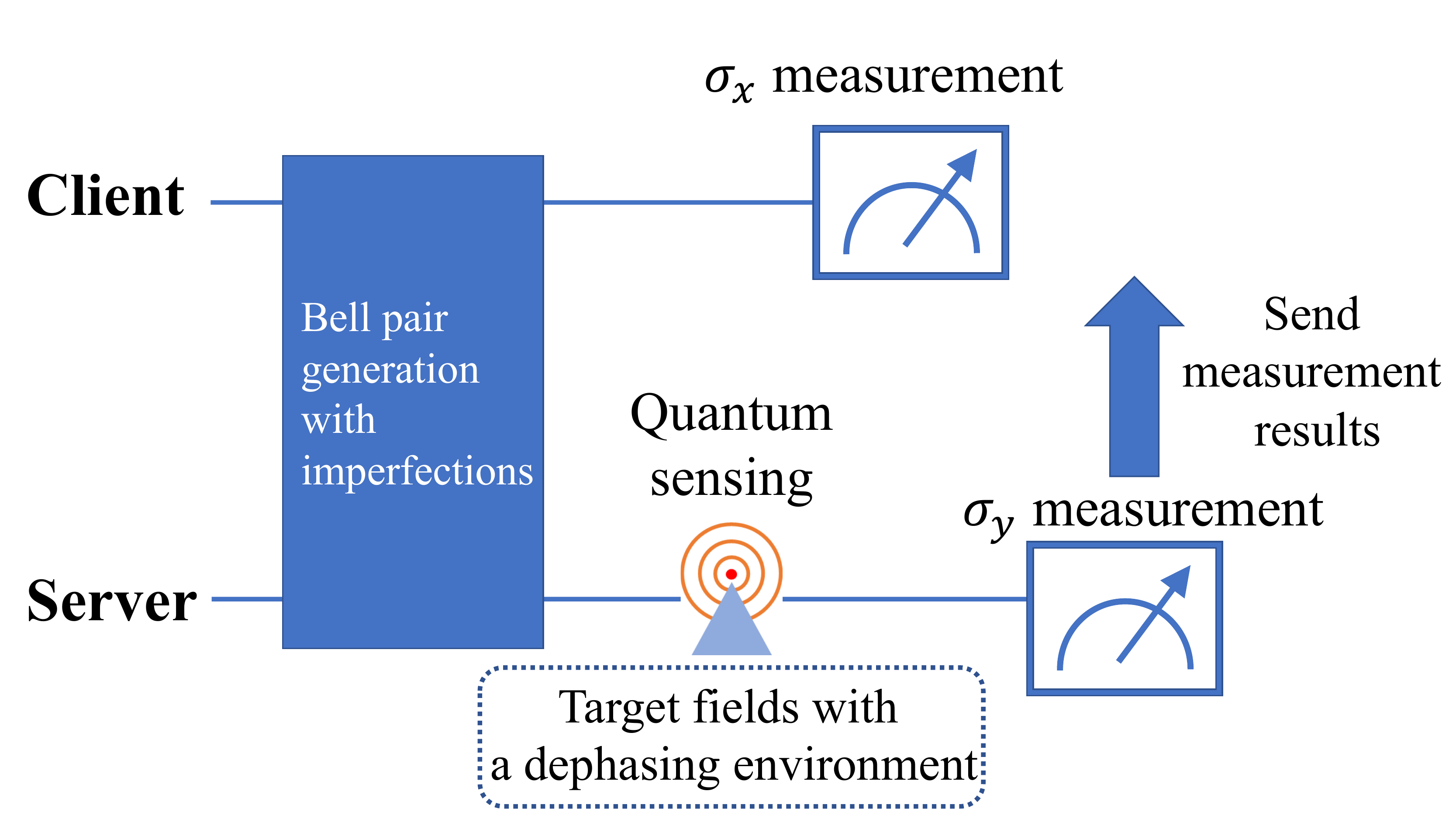}
  \caption{The protocol of the QRS with realistic noise.
  In this protocol, the state shared between the client and the server may deviate from the ideal Bell state, and the fidelity between these two states is evaluated by random-sampling test.
  Also, we consider the effect of dephasing during the quantum sensing.
  }\label{proto}
\end{figure}

In order to investigate realistic situations in our protocol, we should consider both decoherence and state preparation error as shown in Fig.~\ref{proto}, and we introduce these two noise models in this section.
It is worth mentioning that, in the previous research~\cite{takeuchi2019quantum}, only state preparation error was considered.
We assume that the initial state in the server side may deviate from the ideal state $\ket{+}$.
In addition to the imperfect state preparation, we include the effect of dephasing during the interaction with the target magnetic fields, which is considered as one of the major obstacles for quantum metrology.

First, let us explain a noise model to describe dephasing in our protocol \cite{palma1996quantum, de2010universal}.
Our Hamiltonian is written as,
\begin{align}
H(t)&=H_0+H_1(t),\label{ham}\\
H_0&=\frac{\hbar \omega}{2}\sigma_z,\\
H_1(t)&=\hbar\lambda f(t)\sigma_z,
\end{align}
where $\omega$  denotes the qubit frequency, $\lambda$ denotes the coupling strength between the qubit and the environment, and $f(t)$ denotes the classical random variable to express a stochastic noise. 
The Hamiltonian $H_0$ describes the Larmor precession of the qubit to measure the qubit frequency $\omega$. 
It is worth mentioning that there is a linear relationship between the qubit frequency $\omega$ and the target field amplitude $B$ (that the client wants to measure) such as $\omega \propto B$.
This means that the accurate estimation of $\omega $ provides that of the amplitude of the target fields.
The Hamiltonian $H_1$ describes the dephasing of the qubit caused by the environmental noise.
To take into account two typical noise processes, we consider the following correlation functions on the classical random variable $f(t)$,
\begin{align}
 \overline{f(t)}&=0,\label{fbar}\\
 \overline{f(t)f(t')}&=
  \left\{
   \begin{aligned}
    &\tau_{\mathrm{c}}\delta(t-t') &&\text{\ \ white noise}\label{corre}\\
    &1 &&\text{\ \ low frequency noise}
   \end{aligned}
  \right.
\end{align}
where $\tau_{\mathrm{c}}$ and $\delta(t-t')$ denote a correlation time and a Dirac delta function, respectively, and the overline means an ensemble average for the random variable $f(t)$.
In Eqs.~\eqref{fbar} and \eqref{corre}, the noise signal in our model is assumed to be a white noise or a low frequency noise, which frequently appears in the standard quantum metrology.
As shown below, the white noise (low frequency noise) causes the linearly (quadratically) exponential decay in the non-diagonal terms of the density matrix.

Based on our noise model, we solve the time evolution equation for the density operator:
\begin{align}
i\hbar\frac{d\rho(t)}{dt}=-\left[\rho(t),H(t)\right].\label{difeq}
\end{align}
In the interaction picture, Eq.~\eqref{difeq} is rewritten as follows,
\begin{align}
i\hbar\frac{d\rho^{\mathrm{I}}(t)}{dt}&=-\left[\rho^{\mathrm{I}}(t),H_1(t)\right],\label{difeq2}\\
\rho^{\mathrm{I}}(t)&=e^{i H_0 t/\hbar}\rho(t)e^{-i H_0 t/\hbar},
\end{align}
where the initial state is $\rho(0)=\frac{I}{2}+\sum_{i=x, y, z}\frac{r_{i}}{2}\sigma_i$, and $r_i$ denotes the expectation value for the Pauli matrix $\sigma_i$.
By solving Eq.~\eqref{difeq2} and taking the average of the solution, we can get the density matrix with the effect of dephasing,
\begin{align}
 \overline{\rho^{\mathrm{I}}}(t)&=\frac{1+e^{-g(t)}}{2}\rho^{\mathrm{I}}(0)+\frac{1-e^{-g(t)}}{2}\sigma_z\rho^{\mathrm{I}}(0)\sigma_z\label{solrho},\\
 &\quad g(t)=
  \left\{
   \begin{aligned}
    2\lambda^2\tau_{\mathrm{c}}t\ &=t/T_2 &&\text{white noise}\\
    2\lambda^2t^2\ &=(t/T_2)^2 &&\text{low frequency noise}
   \end{aligned}\label{g(t)}
  \right.,
\end{align}
where $g(t)$ is a linear or quadratic function of time for each noise process.
The coupling constant $\lambda$ and the correlation time $\tau_{\mathrm{c}}$ are replaced by the decoherence time $T_2$ for each noise process.
The second term in the right-hand side of Eq.~\eqref{solrho} describes a phase-flip term which decreases the off-diagonal elements of the density matrix.

Second, we assume that the state preparation is not perfect due to errors in a quantum channel between the client and the server.
To evaluate the quality of the state preparation, we utilize the random-sampling test~\cite{nielsen2002quantum,takeuchi2019resource} as with the original QRS~\cite{takeuchi2019quantum} (See also Appendix~\ref{app:b}).
As an advantage of this test compared with the quantum-state tomography, it does not  require any i.i.d.\ property on samples, i.e., it works for any time-varying quantum-channel noise.
Under the success of the random-sampling test, it provides us with a two-qubit state $\rho_{\rm tgt}$ between the client and the server such that
\begin{align}
\braket{\Phi^+|\rho_{\rm tgt}|\Phi^+}&\geq 1-\epsilon \label{fide2}
\end{align}
with high probability, where a finite value $\epsilon$ denotes an error rate that is determined by the number of Bell pairs consumed in the random-sampling test.
Since the client measures his/her part of $\rho_{\rm tgt}$ in the $\sigma_x$ basis, Eq.~\eqref{fide2} means that an initial state $\rho (0)$ such that
\begin{align}
\braket{+|\rho(0)|+}&\geq 1-\epsilon \label{fide}
\end{align}
is prepared at the server's side in the QRS.
Note that, as described in the original QRS~\cite{takeuchi2019quantum}, we can assume that the outcome of the client's $\sigma_x$-basis measurement is always $+1$ that corresponds to the projection onto $|+\rangle$.
If the measurement outcome is $-1$, then the state is supposed to be prepared in $\ket{-}$. However, the client can relabel the basis from $\ket{-}$ to $\ket{+}$. For example, if the client interprets the measurement result of $\sigma_y$ as the other way around such as $+1$ $(-1)$ is interpreted as $-1$ $(+1)$, this is mathematically equivalent to perform $\sigma_z$ operation on the initial state before the state interacts with the magnetic fields, as described in the original QRS~\cite{takeuchi2019quantum}.

\section{\label{sec:uncer}The uncertainty of the estimation under the effect of dephasing}
In this section, we calculate the uncertainty of the estimation under the effect of dephasing.
First, by using the solution of Eq.~\eqref{solrho}, we can derive the uncertainty $\delta\omega_{\mathrm{C}}$ for the client when we perform $\sigma_y$ measurement to extract the information of $\omega$.
The projection operator of $\sigma_y$ measurement in the interaction picture is written as,
\begin{align}
\mathcal{P}^{\mathrm{I}}_y(t)&=(1+\sigma_y^{\mathrm{I}}(t))/2,\label{proje}\\
\sigma_y^{\mathrm{I}}(t)&=e^{i H_0 t/\hbar}\sigma_y e^{-i H_0 t/\hbar}.
\end{align}
We substitute Eq.~\eqref{solrho} and Eq.~\eqref{proje} for the probability $P$,
\begin{align}
P=\mathrm{Tr}\left[\overline{\rho^{\mathrm{I}}}(t)\mathcal{P}^{\mathrm{I}}_y(t)\right]&\simeq x(t)+y(t)\omega t,\label{proby}\\
x(t)=&
\frac{1}{2}+\frac{1}{2}r_ye^{-g(t)},\\
y(t)=&\frac{r_x}{2}e^{-g(t)},
\end{align}
where we assume $\omega t\ll 1$ in the right-hand side of Eq.~\eqref{proby}.
This assumption is valid if the qubit frequency $\omega$ is small and the interaction time $t$, which is typically the order of the decoherence time $T_2$, is short.
Since one of the main purposes of the quantum sensing is to measure small target field, such an assumption is practical for many cases.
We calculate the uncertainty $\delta\omega_{\mathrm{C}}$ ($\delta\omega_{\mathrm{E}}$), which is the uncertainty of the estimation of the target qubit frequency evaluated by the client (Eve).
It is worth mentioning that the client does not know precise form of the initial state.
Based on the assumption that the client only knows the fidelity of the initial state with lack of information about $r_x$, $r_y$, and $r_z$, we can calculate the uncertainty for the client as described in Appendix \ref{app:c},
\begin{align}
\delta\omega_{\mathrm{C}}=\frac{e^{g(t)}}{t\sqrt{M}}\sqrt{1+(M-1)r_y^2 e^{-2g(t)}}.\label{uncerc}
\end{align}
The parameter $r_y$ in Eq.~\eqref{uncerc} originates from the imperfect state preparation and is constrained by the finite fidelity of Eq.~\eqref{fide}.
To consider the worst case, we choose $r_x$ and $r_y$ to maximize $\delta\omega_{\mathrm{C}}$.
Since the uncertainty of Eq.~\eqref{uncerc} is maximized by $r_x=1-2\epsilon$ and $r_y=2\sqrt{\epsilon(1-\epsilon)}$, we obtain the following upper bound for the uncertainty,
\begin{align}
\delta\omega_{\mathrm{C}}\leq\frac{e^{g(t)}}{t\sqrt{M}}\sqrt{1+4(M-1)\epsilon(1-\epsilon) e^{-2g(t)}}\equiv\delta\omega_{\mathrm{C}}^{(U)}\label{uncercm}.
\end{align}
In the expression of $\delta\omega_{\mathrm{C}}^{(U)}$, there is a clear deviation from the central limit theorem that predicts the decrease of $\delta \omega_{\mathrm{C}}^{(U)}$ by $\sqrt{M}$ by increasing the repetition number due to the systematic error for the state preparation.
Actually, for a fixed time $t$, the uncertainty will converges to a non-zero value in the limit of large $M$ ($\delta\omega_{\mathrm{C}}^{(U)}~\rightarrow~ 2\sqrt{\epsilon(1-\epsilon)}/t$).
In the next section, however, we show that, by using the optimized time $t_{\mathrm{opt}}$ to minimize the uncertainty, 
$\delta\omega_{\mathrm{C}}^{(U)}$ slowly converges to zero in the limit of large $M$.

Next, we calculate the uncertainty $\delta\omega_{\mathrm{E}}$ for Eve.
To consider the worst case, we impose two conditions on the calculation of the uncertainty of Eve.
The first condition is that Eve is not affected by dephasing.
This means that Eve can obtain information on the environment and know when the phase-flip error occurs.
So, the time evolution of the density matrix $\rho_{\mathrm{E}}(t)$ of Eve is described only by the free Hamiltonian $H_0$.
The second condition is that Eve knows the information on the error of the state preparation.
This means that the qubit frequency $\omega$ is estimated based on the precise knowledge of the initial state $\rho_E(0)=\frac{I}{2}+\sum_{i=x,y,z}\frac{R_i}{2}\sigma_i$.
The initial state $\rho_E(0)$ is constrained by the following fidelity,
\begin{align}
F(I/2, \rho_E)=\frac{1}{2}+\frac{1}{2}\sqrt{1-R^2}\geq 1-\epsilon,\label{fidee}\\
R^2=R_x^2+R_y^2+R_z^2.
\end{align}
Under the these assumptions, the uncertainty $\delta\omega_{\mathrm{E}}$ is given as (see Appendix \ref{app:c} for detailed calculation),
\begin{align}
\delta\omega_{\mathrm{E}}=\frac{1}{t}\sqrt{\frac{1-R_y^2}{MR_x^2}}.\label{uncere}
\end{align}
Since there is no systematic error from the state preparation in Eq.~\eqref{uncere}, the uncertainty $\delta\omega_{\mathrm{E}}$ decreases by increasing the repetition number $M$, which follows the central limit theorem.
Again, in order to consider the worst case, we minimize the uncertainty $\delta\omega_{\mathrm{E}}$ by choosing $R_x=2\sqrt{\epsilon(1-\epsilon)}$ and $R_y=R_z=0$,
\begin{align}
\delta\omega_{\mathrm{E}}\geq\frac{1}{2t\sqrt{M\epsilon(1-\epsilon)}}\equiv\delta\omega_{\mathrm{E}}^{(L)}.
\end{align}

To compare the amount of the information obtained by the client and Eve, we take the ratio of $\delta\omega_{\mathrm{C}}^{(U)}$ to $\delta\omega_{\mathrm{E}}^{(L)}$,
\begin{align}
 &\delta\omega_{\mathrm{C}}^{(U)}/\delta\omega_{\mathrm{E}}^{(L)}\nonumber\\
 &=2e^{g(t)}\sqrt{\epsilon(1-\epsilon)\left(1+4(M-1)\epsilon(1-\epsilon)e^{-2g(t)}\right)}.\label{ratio}
\end{align}
This ratio is crucial for the QRS.
When this ratio is less than $1$, the client obtains more information than Eve, which is the goal of the QRS.

In the following subsections, we investigate the behavior of the uncertainty $\delta\omega_{\mathrm{C}}^{(U)}$ and the ratio $\delta\omega_{\mathrm{C}}^{(U)}/\delta\omega_{\mathrm{E}}^{(L)}$ in the variation of the repetition number $M$.
The repetition number $M$ is defined as,
\begin{align}
M=\frac{T}{t_{\mathrm{p}}+t+t_{\mathrm{r}}},
\end{align}
where $t_{\mathrm{p}}$ ($t_{\mathrm{r}}$) denotes the preparation (readout) time of the state and $T$ is the total time of the experiment.
If the preparation time and the readout time is much slower than the interaction time ($t_{\mathrm{p}}, t_{\mathrm{r}} \gg t$), we obtain
\begin{align}\label{mtd}
 M\simeq\frac{T}{t_{\mathrm{p}}+t_{\mathrm{r}}}, 
\end{align}
and the repetition number $M$ becomes independent of the interaction time $t$.
On the other hand, if the preparation time and the readout time is much faster than the interaction time ($t_{\mathrm{p}}, t_{\mathrm{r}} \ll t$), the repetition number $M$ can be approximated as
\begin{align}\label{mtind}
M\simeq \frac{T}{t}.
\end{align}
We consider these two cases in the subsections \ref{subsec:slow} and \ref{subsec:fast} of the next section~\ref{sec:opt}.

\section{\label{sec:opt}Optimization of the client's uncertainty}
In order to investigate the condition for the asymmetric gain to be maintained $\delta\omega_{\mathrm{C}}^{(U)}/\delta\omega_{\mathrm{E}}^{(L)}<1$, in this section, we evaluate the uncertainty $\delta\omega_{\mathrm{C}}^{(U)}$ and the ratio $\delta\omega_{\mathrm{C}}^{(U)}/\delta\omega_{\mathrm{E}}^{(L)}$ when the initialization and readout are slow (subsection~\ref{subsec:slow}) and fast (subsection~\ref{subsec:fast}), respectively.
To this end, in each subsection, we optimize the interaction time to minimize the uncertainty $\delta\omega_{\mathrm{C}}^{(U)}$ (and the ratio $\delta\omega_{\mathrm{C}}^{(U)}/\delta\omega_{\mathrm{E}}^{(L)}$ in subsetion~\ref{subsec:fast}).
Table \ref{tab1} shows the table of contents in this section.
\begin{table}[htb]
\begin{tabular}{lccc}
\toprule[1.5pt] &\begin{tabular}{c}Optimization of\\ interaction time $t$\end{tabular}&Optimized $\delta\omega_{\mathrm{C}}$&\begin{tabular}{c}Optimized ratio \\$\delta\omega_{\mathrm{C}}/\delta\omega_{\mathrm{E}}$\end{tabular}\\\hline
Slow & Sec.~V.A.1& Sec.~V.A.2& Sec.~V.A.3 \\
Fast& Sec.~V.B.1& Sec.~V.B.2& Sec.~V.B.3\\\bottomrule[1.5pt]
\end{tabular}
\caption{Table of contents in section V. Slow (Fast) represents slow (fast) initialization and readout.}\label{tab1}
\end{table}

\subsection{\label{subsec:slow}Slow initialization and readout}
\subsubsection{\label{subsubsec:slow}Optimization of the interaction time}
Here,  we calculate the optimized time $t_{\mathrm{opt}}$ to minimize the uncertainty $\delta\omega_{\mathrm{C}}^{(U)}$ in the case of Eq.~\eqref{mtd}, and fix the parameter $\epsilon$ of the state preparation error as $\epsilon=0.001$.
By solving $d(\delta\omega_{\mathrm{C}}^{(U)})/dt=0$ with respect to $t$, we obtain,
\begin{align}
t_{\mathrm{opt}}/T_2=
\left\{
\begin{aligned}
 1+\frac{1}{2}W\left(8(M-1)\epsilon(1-\epsilon)e^{-2}\right)\\
 \text{white noise},\\
 \sqrt{\frac{1+W\left(4(M-1)\epsilon(1-\epsilon)e^{-1}\right)}{2}}\\
 \text{low frequency noise},\label{texact}
 \end{aligned}
\right.
\end{align}
where $W(x)$ is a Lambert W function, by which the inverse solution to $x=ye^y$ is written as $y=W(x)$.
The asymptotic form of the Lambert W function is given as follows,
\begin{align}
 W(x)\simeq
  \left\{
   \begin{aligned}
    &x&&(x\ll 1)\\
    &\ln\left(\frac{x}{\ln(x)}\right)&&(x\gg 1)\label{asympw}
   \end{aligned}
  \right..
\end{align}
This asymptotic form provides an intuition of the asymptotic behavior
of the optimized time $t_{\mathrm{opt}}$, the uncertainty $\delta\omega_{\mathrm{C}}^{(U)}$ and the ratio $\delta\omega_{\mathrm{C}}^{(U)}/\delta\omega_{\mathrm{E}}^{(L)}$.
It is worth mentioning that, depending on the value of $M$ that is a constant determined from experimental conditions, we need to vary the interaction time $t$ for the optimization of the uncertainty.
\begin{figure}
  \centering
  \includegraphics[width=9cm]{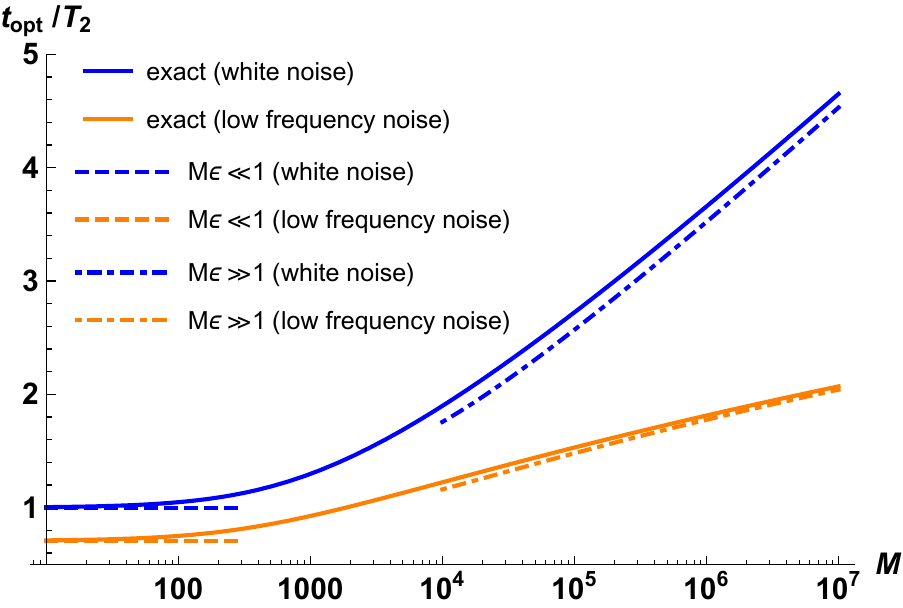}
  \caption{Plots of the optimized time $t_{\mathrm{opt}}$ against the repetition number $M$ with $\epsilon=0.001$ (, which is used for the state preparation error throughout this subsection).
  The solid blue (orange) line denotes the exact solution of the optimized time $t_{\mathrm{opt}}$ in Eq.~\eqref{texact} for the white noise (low frequency noise).
  Also, the approximated form \eqref{toptll} of the optimized time $t_{\mathrm{opt}}$ in the limit of $M\epsilon\ll1$  for the white noise (low frequency noise) is plotted as the dashed blue (orange) line.
Similarly, the approximated form \eqref{toptgg} of the optimized time $t_{\mathrm{opt}}$ in the limit of $M\epsilon\gg1$ for the white noise (low frequency noise) is plotted as the dashed-dotted blue (orange) line.}\label{tplot}
\end{figure}
In Fig.~\ref{tplot}, we show the plot of the optimized time $t_{\mathrm{opt}}$ in terms of $M$.
As the repetition number $M$ with the fixed parameter $\epsilon$ increases, the optimized time $t_{\mathrm{opt}}$ increases, which is different from the standard quantum metrology.
This behavior comes from a competition between the two contributions of dephasing to the uncertainty $\delta\omega_{\mathrm{C}}^{(U)}$ in Eq.~\eqref{uncercm}.
As the first contribution of dephasing, there is an overall factor $e^{g(t)}$ that exponentially increases the uncertainty $\delta\omega_{\mathrm{C}}^{(U)}$ as the interaction time increases, which is typical in the standard quantum metrology.
On the other hand, as the second contribution of dephasing, there is a factor of $\epsilon (1-\epsilon ) e^{-2g(t)}$ that suppresses the systematic error due to the imperfect state preparation in Eq.~\eqref{uncercm}, which appears in our protocol unlike the standard quantum metrology.
Hence, in our protocol, the optimized time $t_{\mathrm{opt}}$ in Eq.~\eqref{texact} is adjusted to control the competition between the two contributions for the minimization of the uncertainty $\delta\omega_{\mathrm{C}}^{(U)}$.

For the regime of $M\epsilon\ll1$, by using the asymptotic form of Eq.~\eqref{asympw}, the optimized time $t_{\mathrm{opt}}$ becomes approximately,
\begin{align}
t_{\mathrm{opt}}/T_2\simeq
\left\{
\begin{aligned}
 1&&\text{white noise},\\
 \frac{1}{\sqrt{2}}&&\text{low frequency noise}.\label{toptll}
 \end{aligned}
\right.
\end{align}
This is the same as the optimized time of the standard quantum metrology introduced in Appendix~\ref{app:a}.
This is because since the term $(M-1)\epsilon(1-\epsilon)$ of the state preparation error in Eq.~\eqref{uncercm} is negligible in the regime of $M\epsilon\ll1$, the uncertainty $\delta\omega_{\mathrm{C}}^{(U)}$ in Eq.~\eqref{uncercm} is reduced to that of the standard quantum metrology.
On the other hand, for the regime of $M\epsilon\gg1$, by using Eq.~\eqref{asympw}, the asymptotic behavior of the optimized time $t_{\mathrm{opt}}$ is written as,
\begin{align}
t_{\mathrm{opt}}/T_2\simeq
\left\{
\begin{aligned}
 \frac{1}{2}\ln\left(\frac{8(M-1)\epsilon(1-\epsilon)}{\ln\left(8(M-1)\epsilon(1-\epsilon)e^{-2}\right)}\right)\\
 \text{white noise},\\
 \sqrt{\frac{1}{2}\ln\left(\frac{4(M-1)\epsilon(1-\epsilon)}{\ln\left(4(M-1)\epsilon(1-\epsilon)e^{-1}\right)}\right)}\\
 \text{low frequency noise}.
 \end{aligned}\label{toptgg}
\right.
\end{align}
The asymptotic behavior of the optimized time $t_{\mathrm{opt}}$ in $M\epsilon\gg1$ is represented as the logarithm of $M\epsilon$.

\subsubsection{Uncertainty of the estimation with the optimized time}
Next, by using the optimized time $t_{\mathrm{opt}}$, we consider the uncertainty $\delta\omega_{\mathrm{C}}^{(U)}$.
By substituting the optimized time $t_{\mathrm{opt}}$ of Eq.~\eqref{texact} for the expression in Eq.~\eqref{uncercm}, the uncertainty $\delta\omega_{\mathrm{C}}^{(U)}$ with the optimized time $t_{\mathrm{opt}}$ can be obtained.
\begin{figure}
  \centering
  \includegraphics[width=9cm]{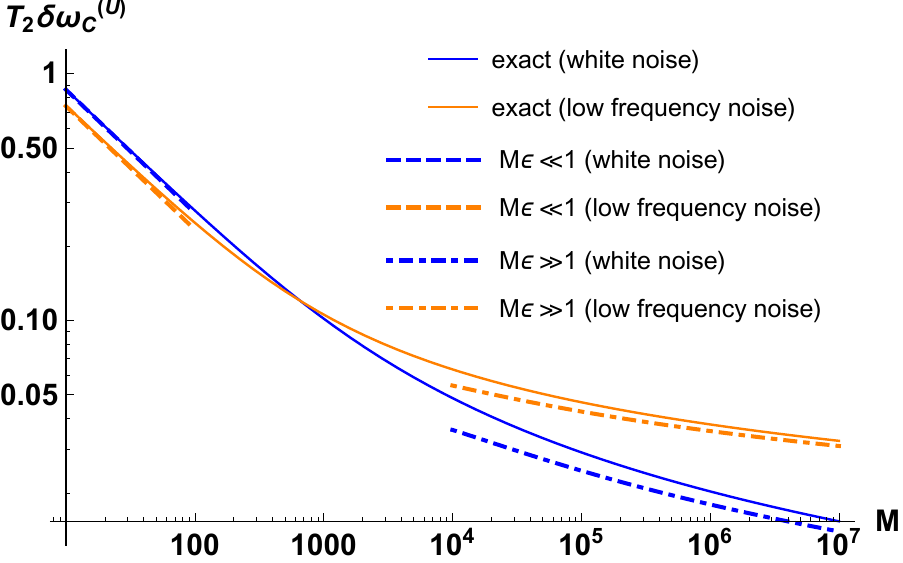}
  \caption{Plots of the uncertainty $\delta\omega_{\mathrm{C}}^{(U)}$ with the optimized time $t_{\mathrm{opt}}$ (that is plotted in Fig.~\ref{tplot}) in terms of the repetition number $M$.
  The solid blue (orange) line denotes the exact uncertainty $\delta\omega_{\mathrm{C}}^{(U)}$ of Eq.~\eqref{uncercm} with the optimized time $t_{\mathrm{opt}}$ of Eq.~\eqref{texact} for the white noise (low frequency noise).
  Also, the approximated uncertainty $\delta\omega_{\mathrm{C}}^{(U)}$ of Eq.~\eqref{uncerll} in the limit of $M\epsilon\ll1$ for the white noise (low frequency noise) is plotted as the dashed blue (orange) line.
  Similarly, the approximated uncertainty $\delta\omega_{\mathrm{C}}^{(U)}$ of Eq.~\eqref{uncergg} in the limit of $M\epsilon\gg1$ for the white noise (low frequency noise) is plotted as the dashed-dotted blue (orange) line.}\label{unplot}
\end{figure}
If the interaction time is determined independently of $M$, the sensitivity would converge to a finite non-zero value in the limit of large $M$ due to the systematic error in the state preparation, as mentioned in Eq.~\eqref{uncercm}.
On the other hand, when we optimize the interaction time, the optimization time logarithmically increases against $M$, as we mentioned.
Such an optimization provides us with a better sensitivity such that the uncertainty of the estimation can asymptotically approach zero by increasing $M$.
Also, in order to investigate the effect of the deviation from the optimized time $t_{\mathrm{opt}}$, we discuss the 
uncertainties 
with the interaction time $t$ around the optimized time $t_{\mathrm{opt}}$ in Appendix~\ref{subsec:slowc}.

In Fig.~\ref{unplot}, the uncertainty $\delta\omega_{\mathrm{C}}^{(U)}$ for the white noise (low frequency noise) is plotted as the solid blue (orange) line with respect to $M$.
In order to understand the behavior of this uncertainty, we use the approximated form of the optimized time in the limit of $M\epsilon\ll1$ and $M\epsilon\gg1$ for the calculation of the approximated uncertainty, and we obtain the following,
\begin{align}
&(M\epsilon\ll1)\nonumber\\
&T_2\delta\omega_{\mathrm{C}}^{(U)}\simeq
\left\{
\begin{aligned}
 \frac{e}{\sqrt{M}}&&\text{white noise},\\
 \sqrt{\frac{2e}{M}}&&\text{low frequency noise},
 \end{aligned}\label{uncerll}
\right.\\\nonumber\\
&(M\epsilon\gg1)\nonumber\\
&T_2\delta\omega_{\mathrm{C}}^{(U)}\simeq
\left\{
\begin{aligned}
 \frac{4\sqrt{\epsilon(1-\epsilon)}}{\ln\left(\frac{8(M-1)\epsilon(1-\epsilon)}{\ln\left(8(M-1)\epsilon(1-\epsilon)e^{-2}\right)}\right)}\\\text{white noise},\\
 2\sqrt{\frac{2\epsilon(1-\epsilon)}{\ln\left(\frac{4(M-1)\epsilon(1-\epsilon)}{\ln\left(4(M-1)\epsilon(1-\epsilon)e^{-1}\right)}\right)}}\\\text{low frequency noise}.
 \end{aligned}\label{uncergg}
\right.
\end{align}
These approximated uncertainties reproduce the behaviors of the exact uncertainties in the regime of either $M\epsilon \ll 1$ or $M\epsilon \gg 1$ as shown in Fig.~\ref{unplot}.
Equation~\eqref{uncerll} shows that the uncertainty decreases in proportion to $1/\sqrt{M}$ as long as the systematic error in the state preparation is negligible, which is the same as the standard quantum metrology introduced in Appendix~\ref{app:a}. 
However, as the repetition number $M$ increases, the solid line of the exact uncertainty deviates from the dashed line of the approximate uncertainty in Eq.~\eqref{uncerll}.
This shows that the effect of the state preparation error in the uncertainty of Eq.~\eqref{uncercm} becomes relevant in the increase of $M$.
It should be noted that the solid blue line and orange line intersect at the point where $M\epsilon\sim \mathcal{O}(1)$ is satisfied.
For the large $M$, the uncertainty $\delta\omega_{\mathrm{C}}^{(U)}$ for the white noise or the low frequency noise is proportional to $1/\ln(M\epsilon)$ or $1/\sqrt{\ln(M\epsilon)}$, respectively, as shown in Eq.~\eqref{uncergg}.
So, in the limit of large $M$, the uncertainty for the white noise decreases faster than that for the low frequency noise.

\subsubsection{Ratio between client's uncertainty and Eve's uncertainty}
In order to investigate how much the information is obtained by Eve, we compare the two uncertainties $\delta\omega_{\mathrm{C}}^{(U)}$ and $\delta\omega_{\mathrm{E}}^{(L)}$ with the optimized time $t_{\mathrm{opt}}$ to minimize the uncertainty $\delta\omega_{\mathrm{C}}^{(U)}$.
By substituting the optimized time $t_{\mathrm{opt}}$ of Eq.~\eqref{texact} for the ratio $\delta\omega_{\mathrm{C}}^{(U)}/\delta\omega_{\mathrm{E}}^{(L)}$ in Eq.\eqref{ratio}, we obtain the ratio with the optimized time.
\begin{figure}
  \centering
  \includegraphics[width=9cm]{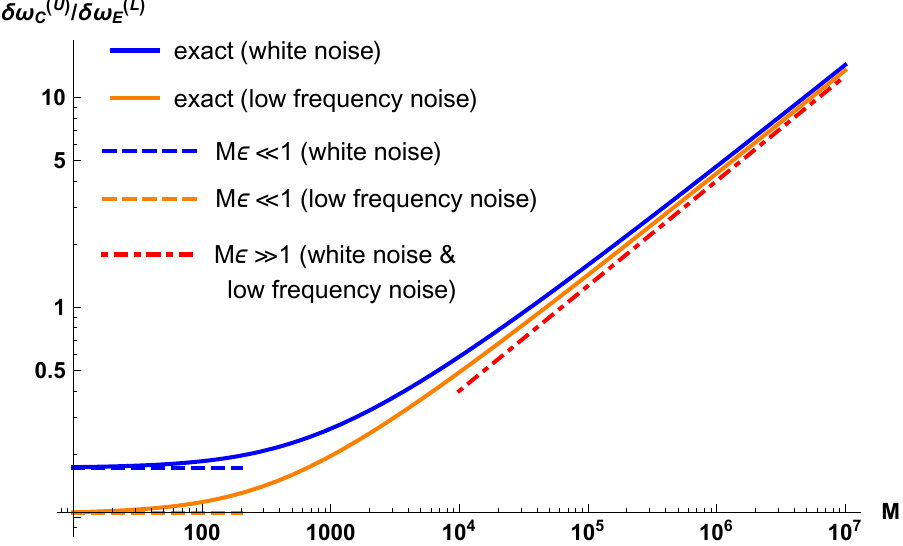}
  \caption{Plots of the ratio $\delta\omega_{\mathrm{C}}^{(U)}/\delta\omega_{\mathrm{E}}^{(L)}$ with the optimized time $t_{\mathrm{opt}}$ (that is plotted in Fig.~\ref{tplot}) in terms of the repetition number $M$.
  The solid blue (orange) line denotes the exact ratio $\delta\omega_{\mathrm{C}}^{(U)}/\delta\omega_{\mathrm{E}}^{(L)}$ of Eq.~\eqref{ratio} with the optimized time $t_{\mathrm{opt}}$ of Eq.~\eqref{texact} for the white noise (low frequency noise).
  Also, the approximated ratio \eqref{ratioll} in the limit of $M\epsilon\ll1$ for the white noise (low frequency noise) is plotted as the dashed blue (orange) line.
  Similarly, the approximated ratio \eqref{ratiogg} in the limit of $M\epsilon\gg1$ for the white noise (low frequency noise) is plotted as the dashed-dotted red line.
  The ratio becomes more than $1$ around when the repetition number $M$ becomes more than $10^5$, 
which shows the conditions for the asymmetric gain $\delta\omega_{\mathrm{C}}/\delta\omega_{\mathrm{E}}<1$
to be maintained.
}\label{raplot}
\end{figure}
In Fig.~\ref{raplot}, we plot the ratio $\delta\omega_{\mathrm{C}}^{(U)}/\delta\omega_{\mathrm{E}}^{(L)}$ for the white noise (low frequency noise) as the solid blue (orange) line in terms of the repetition number $M$.
The ratio becomes more than $1$ around when the repetition number $M$ becomes more than $10^5$.
This means that the client has less information than Eve where the QRS does not provide a suitable asymmetric information gain for the client.

Similarly to the analysis of the uncertainty $\delta\omega_{\mathrm{C}}^{(U)}$, we consider the asymptotic behaviors of the ratio $\delta\omega_{\mathrm{C}}^{(U)}/\delta\omega_{\mathrm{E}}^{(L)}$ in the limit of $M\epsilon\ll1$ and $M\epsilon\gg1$.
By using the asymptotic forms of the optimized time $t_{\mathrm{opt}}$ in Eq.~\eqref{toptll} and Eq.~\eqref{toptgg}, we obtain the approximated ratio $\delta\omega_{\mathrm{C}}^{(U)}/\delta\omega_{\mathrm{E}}^{(L)}$ for the regime of $M\epsilon\ll1$ and $M\epsilon\gg1$ as follows,
\begin{align}
&(M\epsilon\ll1)\nonumber\\
&\delta\omega_{\mathrm{C}}^{(U)}/\delta\omega_{\mathrm{E}}^{(L)}\simeq
\left\{
\begin{aligned}
 2e\sqrt{\epsilon(1-\epsilon)}&&\text{white noise},\\
 2\sqrt{e\epsilon(1-\epsilon)}&&\text{low frequency noise},
 \end{aligned}\label{ratioll}
\right.\\\nonumber\\
&(M\epsilon\gg1)\nonumber\\
&\delta\omega_{\mathrm{C}}^{(U)}/\delta\omega_{\mathrm{E}}^{(L)}\simeq
\left\{
\begin{aligned}
 4\epsilon(1-\epsilon)\sqrt{M}&&\text{white noise},\\
 4\epsilon(1-\epsilon)\sqrt{M}&&\text{low frequency noise}.
 \end{aligned}\label{ratiogg}
\right.
\end{align}
There is a good agreement between the exact solution and approximated solution in Fig.~\ref{raplot}.
As shown in Fig.~\ref{raplot}, the ratio $\delta\omega_{\mathrm{C}}^{(U)}/\delta\omega_{\mathrm{E}}^{(L)}$ is independent of the repetition number for a small $M$ because the state preparation error is negligible. 
As the repetition number increases, the uncertainty of Eve becomes proportional to $1/\sqrt{M}$, while the client can decrease the uncertainty only logarithmically against $M$.
Due to this, the ratio of the uncertainty increases for a larger $M$ as shown in Fig.~\ref{raplot}.
Also, it is worth mentioning that, in the large limit of $M$, the decoherence factor of $g(t)$ disappears from Eq.~\eqref{ratio}, and so the ratio of the uncertainty of the white noise asymptotically approaches the same as that of the low frequency noise.

\subsection{\label{subsec:fast}Fast initialization and readout}
In this subsection, we consider the case of the repetition number $M$ approximated as Eq.~\eqref{mtind} while we set $M$ as the
Eq.~\ref{mtd} in the previous subsection.
The uncertainty $\delta\omega_{\mathrm{C}}^{(U)}$ and the ratio $\delta\omega_{\mathrm{C}}^{(U)}/\delta\omega_{\mathrm{E}}^{(L)}$ are rewritten as follows,
\begin{align}
&\delta\omega_{\mathrm{C}}^{(U)}=\frac{e^{g(t)}}{\sqrt{tT}}\sqrt{1+4\left(T/t-1\right)\epsilon(1-\epsilon) e^{-2g(t)}},\label{uncercut}\\
&\delta\omega_{\mathrm{C}}^{(U)}/\delta\omega_{\mathrm{E}}^{(L)}\nonumber\\
&\qquad=2e^{g(t)}\sqrt{\epsilon(1-\epsilon)\left(1+4\left(T/t-1\right)\epsilon(1-\epsilon)e^{-2g(t)}\right)},\label{ratiot}
\end{align}
where $T$ denotes the total experimental time.
In this subsection, we fix the parameter $\epsilon$ of the state preparation error as $\epsilon=0.0001$ when we plot the figures.

\subsubsection{Optimization of the interaction time}
We can minimize $\delta\omega_{\mathrm{C}}^{(U)}$ of Eq.~\eqref{uncercut} with respect to $t$.
Moreover, we can also minimize the ratio $\delta\omega_{\mathrm{C}}^{(U)}/\delta\omega_{\mathrm{E}}^{(L)}$ as well as the uncertainty $\delta\omega_{\mathrm{C}}^{(U)}$ due to 
the dependence of $M$ on $1/t$, while the ratio $\delta\omega_{\mathrm{C}}^{(U)}/\delta\omega_{\mathrm{E}}^{(L)}$ monotonically
increases against $t$ when $M$ is independent of $t$ as shown in Eq.~\eqref{ratio}.
By solving $\frac{d}{dt}\left(\delta\omega_{\mathrm{C}}^{(U)}\right)=0$ with respect to $t$, we obtain the following equation of the optimized time $t_{\mathrm{opt}}$ for the white noise and the low frequency noise,
\begin{align}
&(2 t_{\mathrm{opt}}/T_2-1)e^{2 t_{\mathrm{opt}}/T_2}\nonumber\\
&\qquad\qquad-4 (1-\epsilon ) \epsilon  \left(2 T/t_{\mathrm{opt}}-1\right)=0\label{toptuw}\\
&\qquad\qquad\qquad\qquad\qquad\qquad\qquad\qquad\text{white noise},\nonumber\\
 &\left(4 \left(t_{\mathrm{opt}}/T_2\right)^2-1\right)e^{2 (t_{\mathrm{opt}}/T_2)^2}\nonumber\\
 &\qquad\qquad-4 (1-\epsilon ) \epsilon  \left(2 T/t_{\mathrm{opt}}-1\right)=0\label{toptul}\\
 &\qquad\qquad\qquad\qquad\qquad\qquad\text{low frequency noise}\nonumber.
\end{align}
Unfortunately, we cannot find an analytical form of the optimized time $t_{\mathrm{opt}}$ to minimize $\delta\omega_{\mathrm{C}}^{(U)}$ in this case, and so we will numerically find the optimized value.
On the other hand, when we try to find the optimized time to minimize the ratio of the uncertainty, we obtain the following analytical solution for the white noise and the low frequency noise by solving $\frac{d}{dt}\left(\delta\omega_{\mathrm{C}}^{(U)}/\delta\omega_{\mathrm{E}}^{(L)}\right)=0$ in terms of $t$
\begin{align}\label{toptr}
t_{\mathrm{optR}}/T_2=
\left\{
\begin{aligned}
 W\left(\sqrt{2 N \epsilon(1-\epsilon )}\right)\\
 \text{white noise},\\
 \frac{\sqrt{3W\left(\frac{4}{3} (N\epsilon(1-\epsilon ) )^{2/3}\right)}}{2}\\
 \text{low frequency noise},
 \end{aligned}
\right.
\end{align}
where $t_{\mathrm{optR}}$ denotes the optimized time for the ratio $\delta\omega_{\mathrm{C}}^{(U)}/\delta\omega_{\mathrm{E}}^{(L)}$ and $N$ is defined as,
\begin{align}
N=T/T_2. \label{defN}
\end{align}

\begin{figure}
  \centering
  \includegraphics[width=9cm]{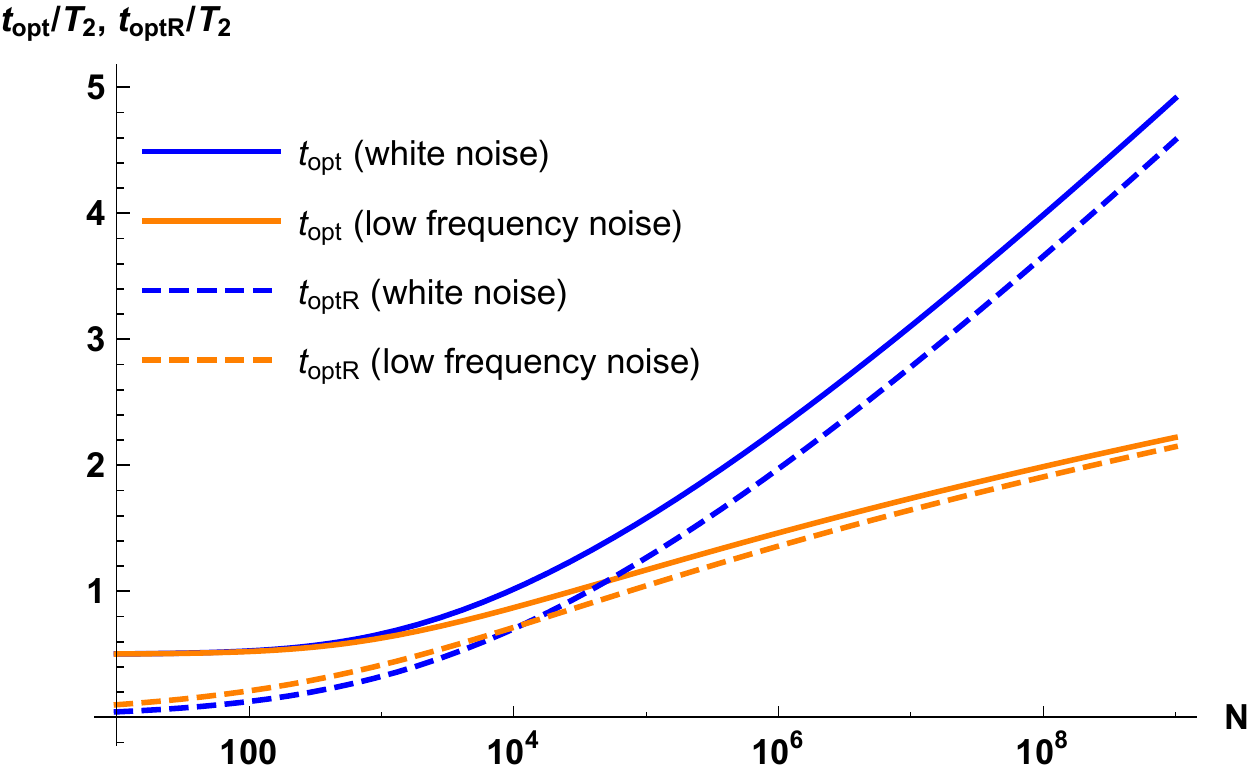}
  \caption{Plots of the optimized time $t_{\mathrm{opt}}$ and $t_{\mathrm{optR}}$ to minimize the uncertainty $\delta\omega_{\mathrm{C}}^{(U)}$ and the ratio $\delta\omega_{\mathrm{C}}^{(U)}/\delta\omega_{\mathrm{E}}^{(L)}$, respectively, with $\epsilon=0.0001$ (, which is used for the state preparation error throughout this subsection).
  The solid blue (orange) line denotes $t_{\mathrm{opt}}$ of Eq.~\eqref{toptuw} (Eq.~\eqref{toptul}) optimized with the uncertainty for the white noise (low frequency noise).
  The dashed blue (orange) line denotes $t_{\mathrm{optR}}$ of Eq.~\eqref{toptr} optimized with the ratio for the white noise (low frequency noise).}\label{totorplot}
\end{figure}
In Fig.~\ref{totorplot}, we plot $t_{\mathrm{opt}}$ of Eq.~\eqref{toptuw} (Eq.~\eqref{toptul}) optimized with the uncertainty $\delta\omega_{\mathrm{C}}^{(U)}$ for the white noise (low frequency noise), and $t_{\mathrm{optR}}$ of 
Eq.~\eqref{toptr} optimized with the ratio $\delta\omega_{\mathrm{C}}^{(U)}/\delta\omega_{\mathrm{E}}^{(L)}$ in terms of $N$.
The both of $t_{\mathrm{opt}}$ and $t_{\mathrm{optR}}$ increase with respect to $N$.
In the limit of $N\epsilon\ll1$, the optimized time $t_{\mathrm{opt}}/T_2$ with the uncertainties for the white noise and the low frequency noise converges to $1/2$, which is the same as the standard quantum metrology in Appendix~\ref{app:a}.
The optimized time $t_{\mathrm{opt}}$ for the white noise is larger than that for the low frequency noise for a finite $\epsilon $, and the difference between them becomes larger as $N$ increases.
On the other hand, the optimized time $t_{\mathrm{optR}}$ for the white noise is smaller than that for the low frequency noise in small $N$.
As $N$ increases, the optimized time $t_{\mathrm{optR}}$ for the white noise overtakes that for the low frequency noise.

\begin{figure*}
  \begin{minipage}[b]{0.45\linewidth}
    \centering
    \includegraphics[width=7.5cm]{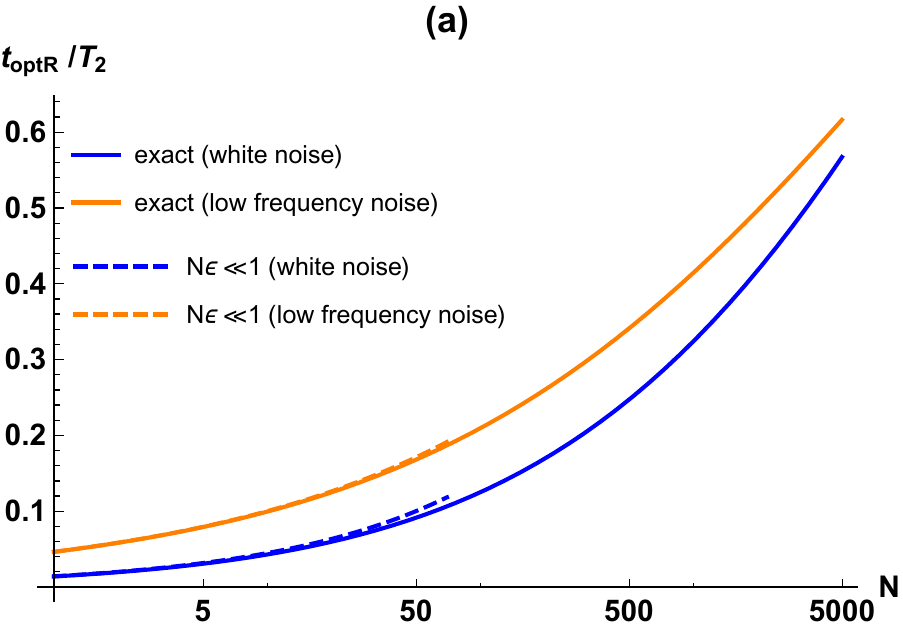}
  \end{minipage}
  \begin{minipage}[b]{0.45\linewidth}
    \centering
    \includegraphics[width=7.5cm]{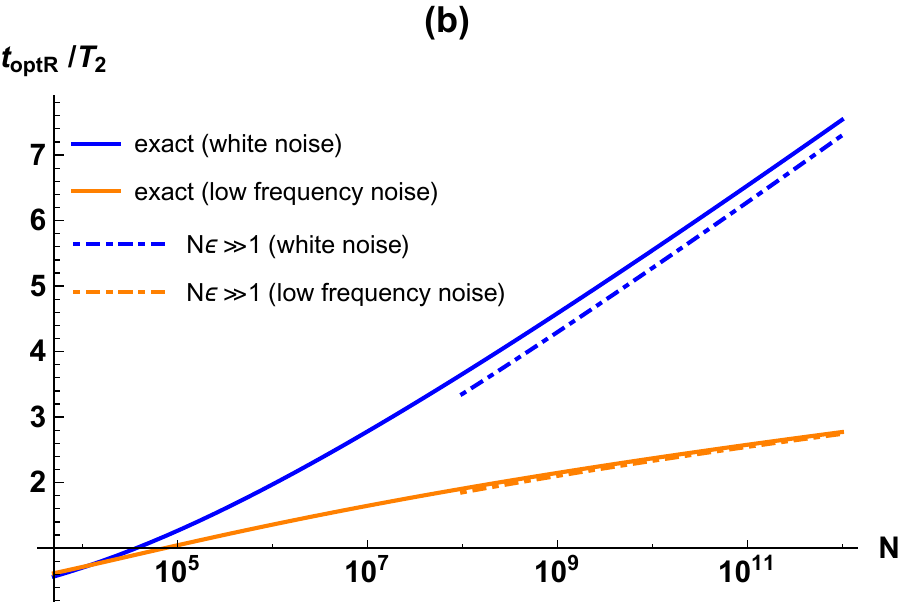}
  \end{minipage}
  \caption{Plots of the optimized time $t_{\mathrm{optR}}$ (that is plotted in Fig.~\ref{totorplot}) for the ratio $\delta\omega_{\mathrm{C}}^{(U)}/\delta\omega_{\mathrm{E}}^{(L)}$. In (a) and (b), the exact solution of the optimized time $t_{\mathrm{optR}}$ in Eq.~\eqref{toptr} for the white noise (low frequency noise) is plotted as the solid blue (orange) line. In (a), the dashed blue (orange) line expresses the optimized time $t_{\mathrm{optR}}$ of Eq.~\eqref{toptrll} approximated in the limit of $N\epsilon\ll1$ for the white noise (low frequency noise). In (b), the dashed-dotted blue (orange) line denotes the optimized time $t_{\mathrm{optR}}$ of Eq.~\eqref{toptrgg} approximated in the limit of $N\epsilon\gg1$ for the white noise (low frequency noise).}\label{torplot}
\end{figure*}

In the limits of $N\epsilon\ll1$ and $N\epsilon\gg1$, we can approximate the optimized time $t_{\mathrm{optR}}$ of Eq.~\eqref{toptr} by using Eq.~\eqref{asympw} as follows,
\begin{align}
 &(N\epsilon\ll1)\nonumber\\
&t_{\mathrm{optR}}/T_2\simeq
\left\{
\begin{aligned}
 \sqrt{2N\epsilon(1-\epsilon)}&&\text{white noise},\\
 \left(N\epsilon(1-\epsilon)\right)^{1/3}&&\text{low frequency noise},
 \end{aligned}\label{toptrll}
\right.\\\nonumber\\
&(N\epsilon\gg1)\nonumber\\
&t_{\mathrm{optR}}/T_2\simeq
\left\{
\begin{aligned}
 \ln\left(\frac{\sqrt{2N\epsilon(1-\epsilon)}}{\ln\left(\sqrt{2N\epsilon(1-\epsilon)}\right)}\right)\\\text{white noise},\\
 \frac{1}{2}\sqrt{3\ln\left(\frac{\frac{4}{3}\left(N\epsilon(1-\epsilon)\right)^{2/3}}{\ln\left(\frac{4}{3}\left(N\epsilon(1-\epsilon)\right)^{2/3}\right)}\right)}\\\text{low frequency noise}.
 \end{aligned}\label{toptrgg}
\right.
\end{align}
In Fig.~\ref{torplot} (a) and (b), we plot the exact solution of the optimized time $t_{\mathrm{optR}}$ for the white noise (low frequency noise) as the solid blue (orange) line.
There is a good agreement between the exact optimized time $t_{\mathrm{optR}}$ and the approximated optimized time $t_{\mathrm{optR}}$ in the regime of $N\epsilon\ll1$ and $N\epsilon\gg1$ in Figs.~\ref{torplot} (a) and (b).

\subsubsection{Uncertainty of the estimation with the optimized time}
Next, we calculate the uncertainty $\delta\omega_{\mathrm{C}}^{(U)}$ with the optimized time $t_{\mathrm{opt}}$ (or $t_{\mathrm{optR}}$) to minimize the uncertainty $\delta\omega_{\mathrm{C}}^{(U)}$ (or the ratio $\delta\omega_{\mathrm{C}}^{(U)}/\delta\omega_{\mathrm{E}}^{(L)}$).
\begin{figure}
  \centering
  \includegraphics[width=9cm]{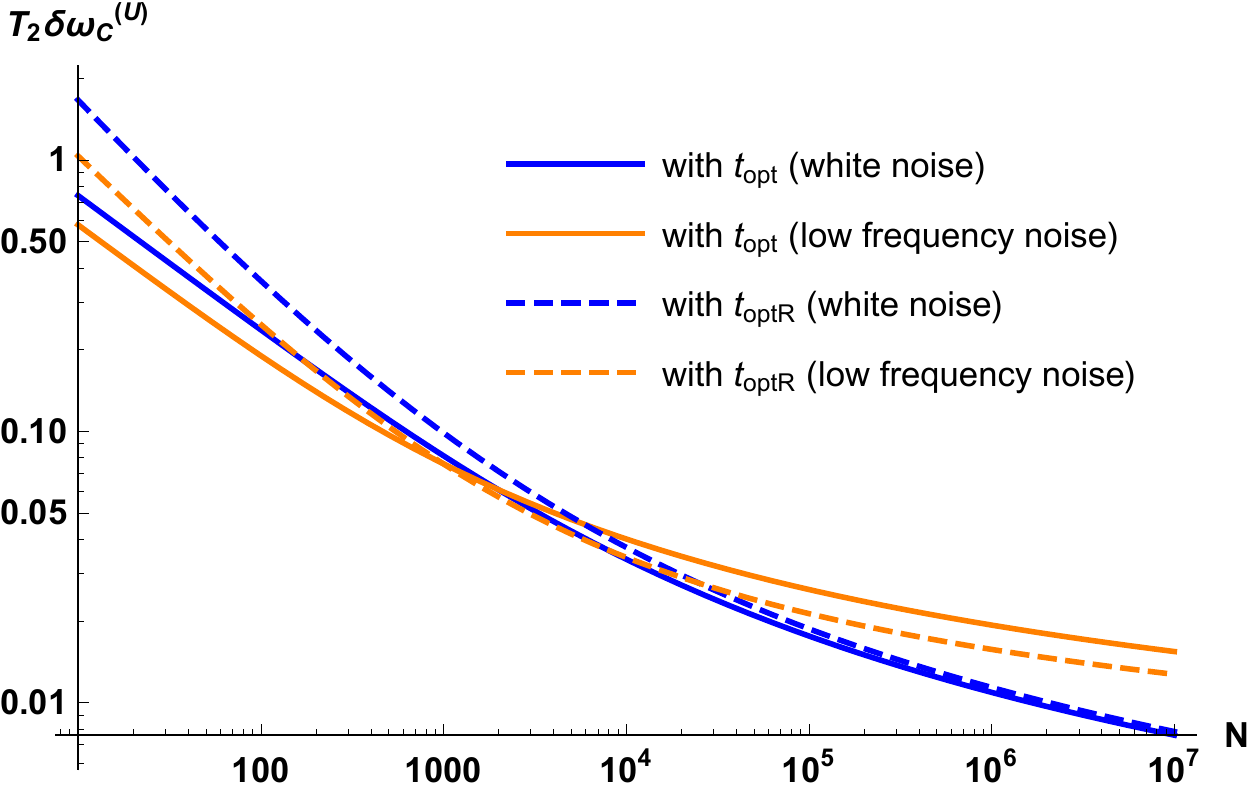}
  \caption{Plots of the uncertainty $\delta\omega_{\mathrm{C}}^{(U)}$ in Eq.~\eqref{uncercut} with the optimized time $t_{\mathrm{opt}}$ ($t_{\mathrm{optR}}$) (that are plotted in Fig.~\ref{totorplot}) to minimize the uncertainty $\delta\omega_{\mathrm{C}}^{(U)}$ (the ratio $\delta\omega_{\mathrm{C}}^{(U)}/\delta\omega_{\mathrm{E}}^{(L)}$).
The solid blue (orange) line denotes the uncertainty minimized by $t_{\mathrm{opt}}$ of Eq.~\eqref{toptuw} (Eq.~\eqref{toptul}) for the white noise (low frequency noise).
The dashed blue (orange) line denotes the uncertainty with $t_{\mathrm{optR}}$ of Eq.~\eqref{toptr} in the optimization of the ratio for the white noise (low frequency noise).}\label{uuroptplot}
\end{figure}
In Fig.~\ref{uuroptplot}, by using the expression of Eq.~\eqref{uncercut}, we plot the uncertainty $\delta\omega_{\mathrm{C}}^{(U)}$ optimized by $t_{\mathrm{opt}}$ of Eq.~\eqref{toptuw} (Eq.~\eqref{toptul}) for the white noise (low frequency noise) as the solid blue (orange) line.
Similarly, in Fig.~\ref{uuroptplot}, the uncertainty $\delta\omega_{\mathrm{C}}^{(U)}$ with $t_{\mathrm{optR}}$ of Eq.~\eqref{toptr} for the white noise (low frequency noise) is plotted as the dashed blue (orange) line.
Fig.~\ref{uuroptplot} shows that the difference between the uncertainties $\delta\omega_{\mathrm{C}}^{(U)}$ with $t_{\mathrm{opt}}$ and $t_{\mathrm{optR}}$ decreases in the increase of $N$.
Furthermore, we can see that the uncertainties $\delta\omega_{\mathrm{C}}^{(U)}$ for the white noise and the low frequency noise intersect at the point where $N\epsilon \sim \mathcal{O}(1)$ both in the optimization with the unceratinty $\delta\omega_{\mathrm{C}}^{(U)}$ and the ratio $\delta\omega_{\mathrm{C}}^{(U)}/\delta\omega_{\mathrm{E}}^{(L)}$.
Also, in order to investigate the effect of the deviation from the optimized time $t_{\mathrm{opt}}$ on the uncertainties, we discuss the uncertainties calculated with the interaction time $t$ around the optimized time $t_{\mathrm{opt}}$ in Appendix~\ref{subsec:fastc}.

In the regime of $N\epsilon\ll1$ and $N\epsilon\gg1$, we analyze the asymptotic behavior of the uncertainty $\delta\omega_{\mathrm{C}}^{(U)}$ with the optimized time $t_{\mathrm{optR}}$ for the ratio $\delta\omega_{\mathrm{C}}^{(U)}/\delta\omega_{\mathrm{E}}^{(L)}$.
By using Eq.~\eqref{asympw}, we can approximate the uncertainty $\delta\omega_{\mathrm{C}}^{(U)}$ with the optimized time $t_{\mathrm{optR}}$ in the limits of $N\epsilon\ll1$ and $N\epsilon\gg1$ as follows,
\begin{align}
 &(N\epsilon\ll1)\nonumber\\
&T_2\delta\omega_C^{(U)}\simeq
\left\{
\begin{aligned}
 \frac{1}{\sqrt{N}}\times\frac{1}{\left(2N\epsilon(1-\epsilon)\right)^{1/4}}\\\text{white noise},\\
 \frac{1}{\sqrt{N}}\times\frac{1}{\left(N\epsilon(1-\epsilon)\right)^{1/6}}\\\text{low frequency noise},
 \end{aligned}\label{urll}
\right.\\\nonumber\\
&(N\epsilon\gg1)\nonumber\\
&T_2\delta\omega_C^{(U)}\simeq
\left\{
\begin{aligned}
 \frac{2\sqrt{\epsilon(1-\epsilon)}}{\ln\left(\frac{\sqrt{2N\epsilon(1-\epsilon)}}{\ln\left(\sqrt{2N\epsilon(1-\epsilon)}\right)}\right)}\\\text{white noise},\\
 4\sqrt{\frac{\epsilon(1-\epsilon)}{3\ln\left(\frac{\frac{4}{3}\left(N\epsilon(1-\epsilon)\right)^{2/3}}{\ln\left(\frac{4}{3}\left(N\epsilon(1-\epsilon)\right)^{2/3}\right)}\right)}}\\\text{low frequency noise}.
 \end{aligned}\label{urgg}
\right.
\end{align}
\begin{figure}
  \centering
  \includegraphics[width=9cm]{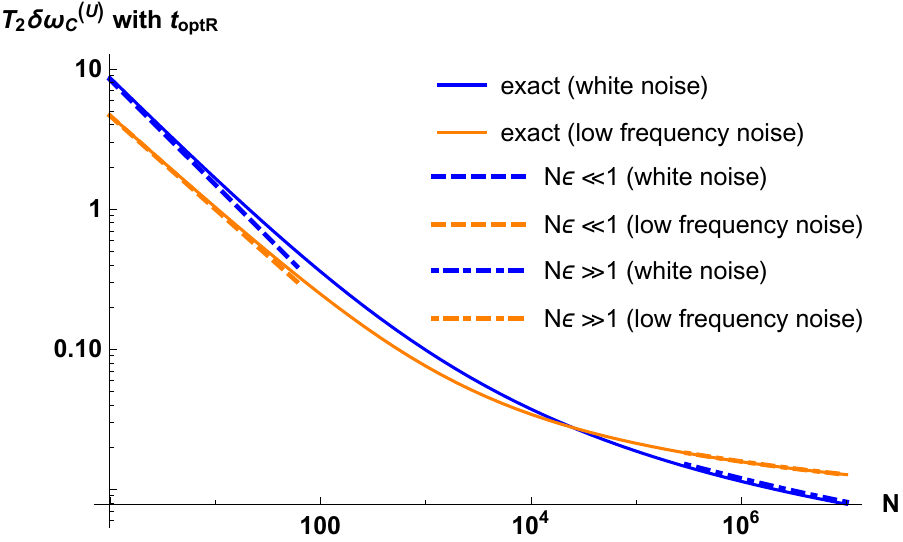}
  \caption{Plots of the exact and approximated uncertainties $\delta\omega_{\mathrm{C}}^{(U)}$ with the optimized time $t_{\mathrm{optR}}$ to minimize the ratio $\delta\omega_{\mathrm{C}}^{(U)}/\delta\omega_{\mathrm{E}}^{(L)}$. 
  The solid blue (orange) line denotes the exact uncertainty $\delta\omega_{\mathrm{C}}^{(U)}$ in Eq.~\eqref{uncercut} with $t_{\mathrm{optR}}$ of Eq.~\eqref{toptr} for the white noise (low frequency noise), which are plotted as the dashed lines in Fig.~\ref{uuroptplot}.
  The dashed blue (orange) line represents the approximated uncertainty $\delta\omega_{\mathrm{C}}^{(U)}$ in Eq.~\eqref{urll} for the white noise (low frequency noise) in the regime of $N\epsilon\ll1$.
  The dashed-dotted blue (orange) line expresses the approximated uncertainty $\delta\omega_{\mathrm{C}}^{(U)}$ in Eq.~\eqref{urgg} for the white noise (low frequency noise) in the regime of $N\epsilon\gg1$.}\label{uroptplot}
\end{figure}
In Fig.~\ref{uroptplot}, the exact uncertainty $\delta\omega_{\mathrm{C}}^{(U)}$ of Eq.~\eqref{uncercut} with $t_{\mathrm{optR}}$ for the white noise (low frequency noise) is plotted as the solid blue (orange) line.
There is a good agreement between the exact and approximated uncertainties $\delta\omega_{\mathrm{C}}^{(U)}$ with $t_{\mathrm{optR}}$ both for the regimes of $N\epsilon\ll1$ and $N\epsilon\gg1$ in Fig.~\ref{uroptplot}.
The uncertainty $\delta\omega_{\mathrm{C}}^{(U)}$ of Eq.~\eqref{urll} approximated in small $N\epsilon$ for the white noise and the low frequency noise decreases in proportion to $N^{-3/4}$ and $N^{-2/3}$, respectively.
It is worth mentioning that the uncertainty seems to beat the classical scaling of $\delta \omega_{\mathrm{C}}^{(U)} \propto N^{-1/2}$, which comes from the central limit theorem.
However, such a scaling of $N^{-3/4}$ or $N^{-2/3}$ holds only when $N\epsilon$ is much smaller than 1.
Actually, once $N\epsilon$ becomes much larger than 1, the uncertainty decreases logarithmically, which is below the classical scaling.
We also observe that there is the intersection between the exact uncertainties $\delta\omega_{\mathrm{C}}^{(U)}$ for the white noise and the low frequency noise in the intermediate region of $N\epsilon$.

\subsubsection{Ratio between client's uncertainty and Eve's uncertainty}
Here, we discuss how the behavior of the ratio $\delta\omega_{\mathrm{C}}^{(U)}/\delta\omega_{\mathrm{E}}^{(L)}$ depends on the choice of $t_{\mathrm{opt}}$ and $t_{\mathrm{optR}}$.
\begin{figure}
  \centering
  \includegraphics[width=9cm]{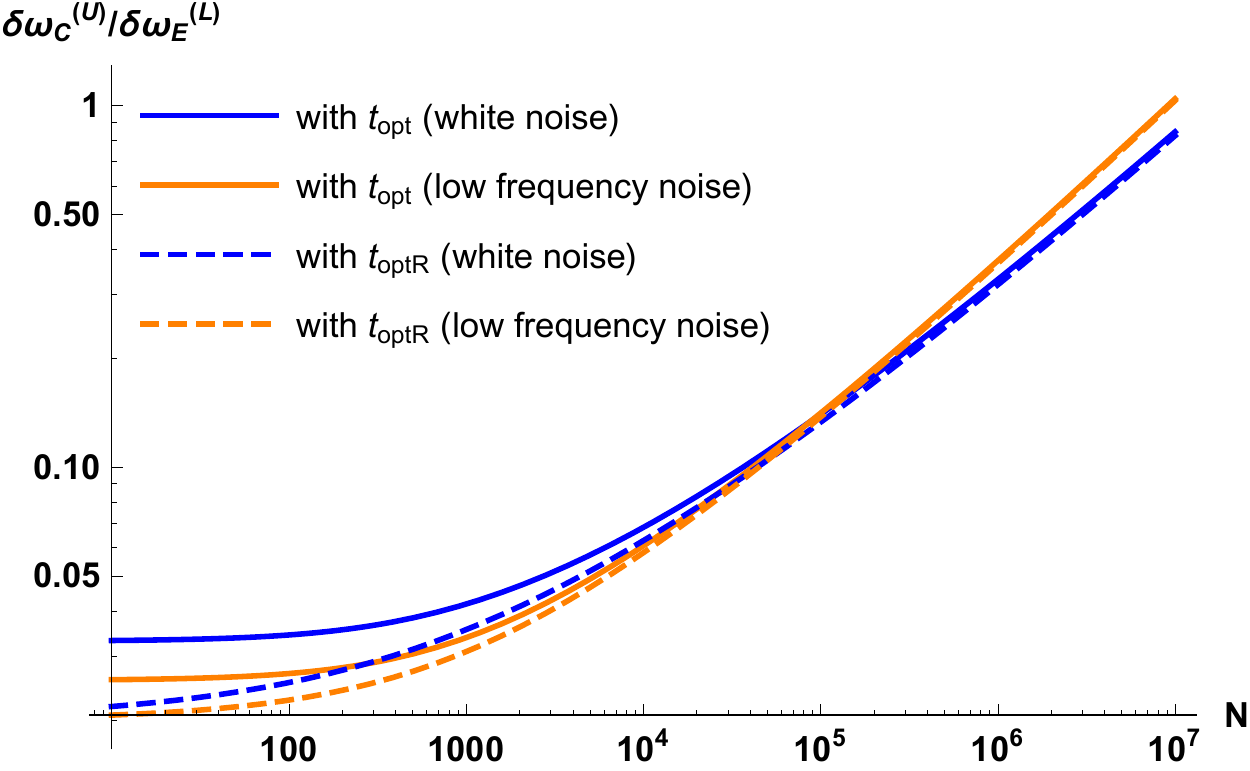}
  \caption{Plots of the ratio $\delta\omega_{\mathrm{C}}^{(U)}/\delta\omega_{\mathrm{E}}^{(L)}$ in Eq.~\eqref{ratiot} with the optimized time $t_{\mathrm{opt}}$ ($t_{\mathrm{optR}}$) (that are plotted in Fig.~\ref{totorplot}) to minimize the uncertainty $\delta\omega_{\mathrm{C}}^{(U)}$ (the ratio $\delta\omega_{\mathrm{C}}^{(U)}/\delta\omega_{\mathrm{E}}^{(L)}$).
The solid blue (orange) line denotes the ratio with $t_{\mathrm{opt}}$ of Eq.~\eqref{toptuw} (Eq.~\eqref{toptul}) in the optimization of the client's uncertainty for the white noise (low frequency noise).
The dashed blue (orange) line denotes the ratio minimized by $t_{\mathrm{optR}}$ of Eq.~\eqref{toptr} for the white noise (low frequency noise).
The ratio becomes more than $1$ around when $N$ becomes more than $10^7$.}\label{rrroptplot}
\end{figure}
The plot of Fig.~\ref{rrroptplot} shows the ratio $\delta\omega_{\mathrm{C}}^{(U)}/\delta\omega_{\mathrm{E}}^{(L)}$ with the optimized time $t_{\mathrm{opt}}$ ($t_{\mathrm{optR}}$).
In Fig.~\ref{rrroptplot}, as $N$ becomes larger, the differences between the ratio $\delta\omega_{\mathrm{C}}^{(U)}/\delta\omega_{\mathrm{E}}^{(L)}$ with $t_{\mathrm{opt}}$ and that with $t_{\mathrm{optR}}$ become smaller for both the white noise and the low frequency noise.
Moreover, regardless of whether we optimize the uncertainty $\delta\omega_{\mathrm{C}}^{(U)}$ or the ratio $\delta\omega_{\mathrm{C}}^{(U)}/\delta\omega_{\mathrm{E}}^{(L)}$, the ratios $\delta\omega_{\mathrm{C}}^{(U)}/\delta\omega_{\mathrm{E}}^{(L)}$ of the white noise and the low frequency noise intersect at the point where $N\epsilon$ is around one to ten.
The ratios become more than $1$ around when the repetition number $M$ becomes more than $10^7$, and this means that the client has less information than Eve where the QRS does not provide a suitable asymmetric information gain for the client.
Also, in Appendix~\ref{subsec:fastc}, in order to investigate the effect of the deviation from the optimized time $t_{\mathrm{optR}}$, we discuss the ratios with the interaction time $t$ around the optimized time $t_{\mathrm{optR}}$.

Based on the analytical solution of the optimized time $t_{\mathrm{optR}}$ in Eq.~\eqref{toptr}, we investigate the asymptotic behavior of the ratio $\delta\omega_{\mathrm{C}}^{(U)}/\delta\omega_{\mathrm{E}}^{(L)}$ with the optimized time in the regime of $N\epsilon\ll1$ and $N\epsilon\gg1$.
By substituting the approximation for the Lambert W function in Eq.~\eqref{asympw} into the ratio $\delta\omega_{\mathrm{C}}^{(U)}/\delta\omega_{\mathrm{E}}^{(L)}$ in Eq.~\eqref{ratiot}, the asymptotic form of the ratio can be approximated as follows,
\begin{align}
 &(N\epsilon\ll1)\nonumber\\
&\delta\omega_{\mathrm{C}}^{(U)}/\delta\omega_{\mathrm{E}}^{(L)}\simeq
\left\{
\begin{aligned}
 2\sqrt{\epsilon(1-\epsilon)}&&\text{white noise},\\
 2\sqrt{\epsilon(1-\epsilon)}&&\text{low frequency noise},
 \end{aligned}\label{rll}
\right.\\\nonumber\\
&(N\epsilon\gg1)\nonumber\\
&\delta\omega_{\mathrm{C}}^{(U)}/\delta\omega_{\mathrm{E}}^{(L)}\simeq
\left\{
\begin{aligned}
 4\epsilon(1-\epsilon)\sqrt{\frac{N}{\ln\left(\frac{\sqrt{2N\epsilon(1-\epsilon)}}{\ln\left(\sqrt{2N\epsilon(1-\epsilon)}\right)}\right)}}\\\text{white noise},\\
 4\epsilon(1-\epsilon)\sqrt{\frac{2N}{\sqrt{3}\ln\left(\frac{\frac{4}{3}\left(N\epsilon(1-\epsilon)\right)^{2/3}}{\ln\left(\frac{4}{3}\left(N\epsilon(1-\epsilon)\right)^{2/3}\right)}\right)}}\\\text{low frequency noise}.
 \end{aligned}\label{rgg}
\right.
\end{align}
\begin{figure}
  \centering
  \includegraphics[width=9cm]{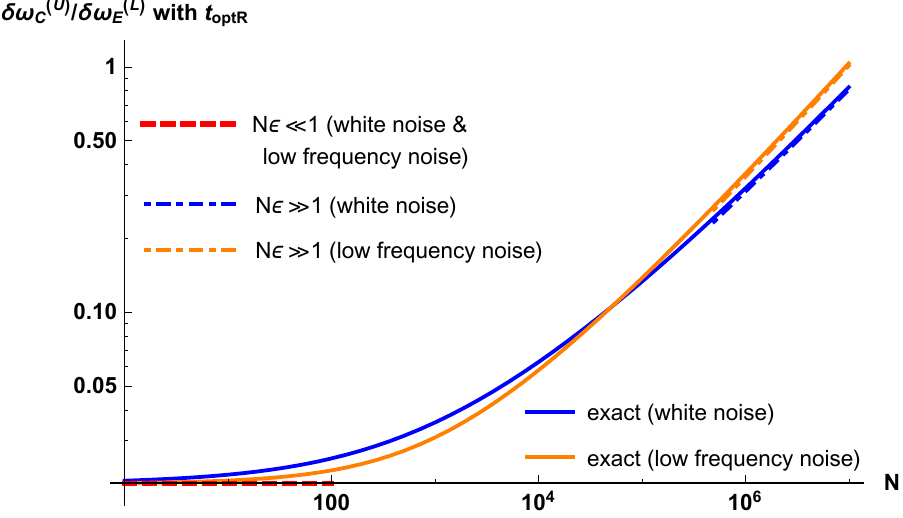}
  \caption{Plots of the exact and approximated ratios $\delta\omega_{\mathrm{C}}^{(U)}/\delta\omega_{\mathrm{E}}^{(L)}$ with the optimized time $t_{\mathrm{optR}}$ to minimize the ratio $\delta\omega_{\mathrm{C}}^{(U)}/\delta\omega_{\mathrm{E}}^{(L)}$. 
  The solid blue (orange) line denotes the exact ratio $\delta\omega_{\mathrm{C}}^{(U)}/\delta\omega_{\mathrm{E}}^{(L)}$ in Eq.~\eqref{ratiot} with the optimized time $t_{\mathrm{optR}}$ of Eq.~\eqref{toptr} for the white noise (low frequency noise), which are plotted as the dashed lines in Fig.~\ref{rrroptplot}.
  The dashed blue (orange) line represents the approximated ratio $\delta\omega_{\mathrm{C}}^{(U)}/\delta\omega_{\mathrm{E}}^{(L)}$ of Eq.~\eqref{rll} for the white noise (low frequency noise) in the regime of $N\epsilon\ll1$.
  The dashed-dotted blue (orange) line expresses the approximated ratio $\delta\omega_{\mathrm{C}}^{(U)}/\delta\omega_{\mathrm{E}}^{(L)}$ in Eq.~\eqref{rgg} for the white noise (low frequency noise) in the regime of $N\epsilon\gg1$.}\label{rroptplot}
\end{figure}\noindent
In Fig.~\ref{rroptplot}, we plot the exact ratio $\delta\omega_{\mathrm{C}}^{(U)}/\delta\omega_{\mathrm{E}}^{(L)}$ with the optimized time $t_{\mathrm{optR}}$ as the solid blue (orange) line for the white noise (low frequency noise).
As seen from Fig.~\ref{rroptplot}, the approximate ratios of $\delta\omega_{\mathrm{C}}^{(U)}/\delta\omega_{\mathrm{E}}^{(L)}$ reproduce the asymptotic form of the exact ratio both in the regime of $N\epsilon\ll1$ and $N\epsilon\gg1$.

\section{\label{sec:Concl}Summary and Conclusion}
In conclusion we have investigated the effect of dephasing for the QRS protocol.
The original paper~\cite{takeuchi2019quantum} on the QRS considers the state preparation error caused by the channel noise between the client and the server, and evaluates the fidelity of the shared state by using the random-sampling test.
In addition to the state preparation error, we introduce the dephasing during the quantum sensing, which is one of the most typical noise for solid state systems. 
We show that the uncertainty of the client side decreases with the square root of the repetition number $M$ for
small $M$.  On the other hand, for
large $M$, the state preparation error becomes as relevant as the dephasing, and the uncertainty $\delta\omega_{\mathrm{C}}$ for the client side
decreases logarithmically with $M$. This is the nontrivial result in our paper 
because the uncertainty decreases with the square root of $M$ in the standard quantum metrology with the perfect state preparation.
Moreover, we calculate the uncertainty of the qubit frequency $\delta\omega_{\mathrm{E}}$ for the Eve, and compare this with  $\delta\omega_{\mathrm{C}}$.
Our results lead us to obtain the conditions for the asymmetric gain $\delta\omega_{\mathrm{C}}/\delta\omega_{\mathrm{E}}<1$ to be maintained even under the effect of dephasing, which is an important step for the realization of the quantum remote sensing with solid state systems.
The results to obtain this condition are shown in Table~\ref{tab2} and \ref{tab3}.
\begin{center}
\begin{table*}[ht]
{\small
\hfill{}
\begin{tabular}{lcc}
\toprule[1.5pt] &$t_{\mathrm{opt}}$&$\delta\omega_{\mathrm{C}}^{(U)}/\delta\omega_{\mathrm{E}}^{(L)}$\\\hline\addlinespace[2pt]
Slow ~&  $\begin{aligned}&1+W\left(8(M-1)\epsilon(1-\epsilon)e^{-2}\right)/2 && \text{(white)}\\&\sqrt{(1+W\left(4(M-1)\epsilon(1-\epsilon)e^{-1}\right))/2}&&\text{(low)}\end{aligned}$&~~$\begin{aligned}&\Bigl[4\epsilon(1-\epsilon)\times\Bigl(e^{2g(t_{\mathrm{opt}})}\Bigr.\Bigr.&&\\&\Bigl.\Bigl.\qquad+4(M-1)\epsilon(1-\epsilon)\Bigr)\Bigr]^{1/2}&&\end{aligned}$ \\
\hline\addlinespace[1.5pt]
Fast & \multicolumn{2}{c}{Numerical calculation (shown in Figs.~\ref{totorplot} and \ref{rrroptplot})} \\\bottomrule[1.5pt]
\end{tabular}}
\hfill{}
\caption{Summary of our results. The optimized interaction time $t_{\mathrm{opt}}$ and a ratio of the uncertainty $\delta\omega_{\mathrm{C}}^{(U)}/\delta\omega_{\mathrm{E}}^{(L)}$ where we perform the optimization to minimize $\delta\omega_{\mathrm{C}}^{(U)}$.
Here, $\delta\omega_{\mathrm{C}}^{(U)}$ ($\delta\omega_{\mathrm{E}}^{(L)}$) shows an upper (lower) bound of the uncertainty of the Client (Eve), $\epsilon $ shows a state preparation error, $M$ denotes the repetition number, $g(t)$ denotes a dephasing effect coming from either white or low frequency noise. 
We consider a case where the readout and initialization is much faster or slower than the interaction time with the target fields.}\label{tab2}
\end{table*}
\end{center}
\begin{center}
\begin{table*}[ht]
{\small
\hfill{}
\begin{tabular}{lcc}
\toprule[1.5pt] &$t_{\mathrm{optR}}$&$\delta\omega_{\mathrm{C}}^{(U)}/\delta\omega_{\mathrm{E}}^{(L)}$\\\hline\addlinespace[2pt]
Slow & \multicolumn{2}{c}{Do not exist} \\\hline\addlinespace[1.5pt]
Fast ~& $\begin{aligned}&W(\sqrt{2 N \epsilon(1-\epsilon )})&& \text{(white)}\\&\sqrt{3W\left(4/3 (N\epsilon(1-\epsilon ) )^{2/3}\right)}/2&&\text{(low)}\end{aligned}$
~~ &~~$\begin{aligned}&\Bigl[4\epsilon(1-\epsilon)\times\Bigl(e^{2g(t_{\mathrm{optR}})}\Bigr.\Bigr.&&\\&\Bigl.\Bigl.\qquad+4(NT_2/t_{\mathrm{optR}}-1)\epsilon(1-\epsilon)\Bigr)\Bigr]^{1/2}&&\end{aligned}$
\\\bottomrule[1.5pt]
\end{tabular}}
\hfill{}
\caption{Summary of our results. The optimized interaction time $t_{\mathrm{optR}}$ and a ratio of the uncertainty $\delta\omega_{\mathrm{C}}^{(U)}/\delta\omega_{\mathrm{E}}^{(L)}$ where we perform the optimization to minimize $\delta\omega_{\mathrm{C}}^{(U)}/\delta\omega_{\mathrm{E}}^{(L)}$.
When the readout and initialization is much slower than the interaction time, $\delta\omega_{\mathrm{C}}^{(U)}/\delta\omega_{\mathrm{E}}^{(L)}$ monotonically increases against the interaction time, and so there is no optimal time.
We use the same notation as that used in the Table~\ref{tab2}.}\label{tab3}
\end{table*}
\end{center}

\section*{Acknowledgement}
We are grateful to Shiro Kawabata for useful discussion.
This work was supported by Leading Initiative for Excellent Young Researchers MEXT Japan and JST presto (Grant No. JPMJPR1919) Japan.
This work was also supported by CREST (JPMJCR1774), JST.


\appendix
\section{\label{app:a}Standard quantum metrology}
We review the standard quantum metrology implemented under ideal conditions.
The Hamiltonian is given as,
\begin{align}
H_0=\frac{\hbar\omega}{2}\sigma_z \label{ham0},
\end{align}
where the frequency of qubit $\omega$ has a linear relationship with the amplitude of the target field $B$ that we want to know, and we  will assume $\omega t\ll1$ for the weak target field $B$.
The steps of the standard quantum metrology are as follows:
\begin{enumerate}
\item An initial state $\ket{+}$ is prepared.
\item The state $\ket{+}$ is evolved with the Hamiltonian in Eq.~\eqref{ham0} for an interaction time $t$.
\item The $\sigma_y$ measurement is performed for the final state.
\item Repeat the steps 1 - 3, $M$ times.
\end{enumerate}
In the step 2, the state of the qubit acquires a relative phase due to the interaction between the qubit and the target magnetic field.
By performing the $\sigma_y$ measurement in the step 3, the relative phase acquired by the target field is measured.
In the actual experiment, the $\sigma_y$ measurement in the step 3 produces the outcome $m(=1~\rm{or}~ 0)$ where $m=1~(0)$ corresponds to the eigenvalue $+1~(-1)$ for $\sigma_y$.
The $\sigma_y$ measurement is repeated $M$ times and the average value $S_M$ is defined as
\begin{align}
S_M=\frac{\sum_{j=1}^M m_j}{M},
\end{align}
where $m_j$ denotes the outcome of the $j$-th $\sigma_y$ measurement.

We can calculate the probability $P$ of the outcome $m=1$ for the $\sigma_y$ measurement as follows,
\begin{align}
P&=\mathrm{Tr}\left[\mathcal{P}_y e^{-iH_0 t/\hbar}\ket{+}\bra{+}e^{iH_0 t/\hbar}\right]\\
&=(1+\sin{\omega t})/2\\
&\simeq (1+\omega t)/2\label{proba}
\end{align}
where $\mathcal{P}_y=(1+\sigma_y)/2$ denotes the projection operator for the $\sigma_y$ measurement, and we use the approximation of $\omega t\ll1$ in the last line.
By replacing the probability $P$ in Eq.~\ref{proba} with the average value $S_M$, the frequency $\omega$ can be estimated as,
\begin{align}
\omega_M^{(\mathrm{est})}=(2S_M-1)/t. \label{estuna}
\end{align}
The uncertainty $\delta\omega$ of the estimation is defined as the root mean squared error,
\begin{align}
\delta\omega&\equiv \sqrt{\braket{(\omega^{\mathrm{(est)}}_M-\omega)^2}}\label{defun}\\
&=\frac{2}{t}\sqrt{\braket{(S_M-P)^2}},
\end{align}
where we substitute the estimation $\omega^{\mathrm{(est)}}_M$ of Eq.~\eqref{estuna} and the frequency $\omega$ of Eq.~\eqref{proba} for the uncertainty $\delta\omega$.
Since the relation between the variance of the average value $S_M$ and  the variance $\delta^2P$ of the random variable $m$ is given as follows,
\begin{align}
M\braket{(S_M-P)^2}=\delta^2P,
\end{align}
where $\delta^2P=P(1-P)\simeq 1/4$ under the assumption of $\omega t\ll1$, we obtain the uncertainty $\delta\omega$ of the estimation,
\begin{align}
\delta\omega\simeq \frac{1}{t\sqrt{M}}.
\end{align}

While we have considered a quantum sensing without dephasing, we will explain the case with dephasing below.
By replacing the probability $P$ in Eq.~\eqref{proba} with the new probability $P$ including the effect of dephasing as follows,
\begin{align}
P\simeq(1+e^{-g(t)}\omega t)/2, \label{probade}
\end{align}
we obtain the uncertainty with dephasing as
\begin{align}
\delta\omega \simeq \frac{e^{g(t)}}{t\sqrt{M}},\label{stunde}
\end{align}
where the factor $e^{g(t)}$ expresses the loss of quantum coherence for the qubit, and $g(t)$ is a linear or quadratic function of the interaction time $t$ as defined in Eq.~\eqref{g(t)}.
By calculating $\frac{d}{dt}\left(\delta\omega\right)=0$, the optimized time $t_{\mathrm{opt}}$ to minimize the uncertainty of Eq.~\eqref{stunde} is given as follows,
\begin{align}
t_{\mathrm{opt}}/T_2=
\left\{
\begin{aligned}
 1&&\text{white noise},\\
 \frac{1}{\sqrt{2}}&&\text{low frequency noise},\label{stoptt}
 \end{aligned}
\right.
\end{align}
where $T_2$ denotes the decoherence time.
By using the optimized time $t_{\mathrm{opt}}$, the uncertainty can be obtained as,
\begin{align}
\delta\omega=
\left\{
\begin{aligned}
 \frac{1}{T_2}\frac{e}{\sqrt{M}}&&\text{white noise},\\
 \frac{1}{T_2}\sqrt{\frac{2e}{M}}&&\text{low frequency noise}.
 \end{aligned}\label{stunopt}
\right.
\end{align}
When the time of the initialization and the readout of the qubit are much shorter than the interaction time $t$, the repetition number $M$ is written in terms of the total time $T$ of experiment and the interaction time $t$,
\begin{align}
M=\frac{T}{t}.
\end{align}
In this case, the optimized time $t_{\mathrm{opt}}$ is calculated as
\begin{align}
t_{\mathrm{opt}}/T_2=
\left\{
\begin{aligned}
 \frac{1}{2}&&\text{white noise},\\
 \frac{1}{2}&&\text{low frequency noise}.\label{stopttf}
 \end{aligned}
\right.
\end{align}
By using these optimized time $t_{\mathrm{opt}}$, the uncertainty is obtained as follows,
\begin{align}
\delta\omega=
\left\{
\begin{aligned}
 \sqrt{\frac{2e}{T_2 T}}&&\text{white noise},\\
 \sqrt{\frac{2\sqrt{e}}{T_2 T}}&&\text{low frequency noise}.\label{stunoptf}
 \end{aligned}
\right.
\end{align}
The results of Eqs.~\eqref{probade} - \eqref{stunoptf} are known as the results of standard quantum metrology under the effect of dephasing.

\section{\label{app:b}Random-sampling test}
We review the concept of the random-sampling test.
A key point is to measure two stabilizer operators $\sigma_x\otimes\sigma_x$ and $\sigma_z\otimes\sigma_z$, which gives us a lower bound on the fidelity.
If a measured state is the ideal Bell state $\ket{\Phi^+}$, the outcomes of these two stabilizer measurements are always $+1$, which means that the measurements on the first and the second qubits return the same outcomes.
First, the server prepares an $8k$-qubit state $\rho_S$, where $k=\lceil75\ln{(2/\delta)}/[8(\epsilon-3\Delta)^2]\rceil$.
Without loss of generality, we consider that $\rho_S$ consists of $4k$ registers, and each register stores two qubits.
Although the state $\rho_S$ of $4k$ registers is $|\Phi^+\rangle^{\otimes 4k}$ in the ideal case, it is arbitrary $8k$-qubit state when the quantum channel is noisy.
Second, the server sends the client one half of each register one by one.
Then the client chooses $2k$ registers among $4k$ registers independently and uniformly at random.
For the first $k$ registers, the client and the server measure their own half in the $\sigma_x$ basis.
For another $k$ registers, they measure their halves in the $\sigma_z$ basis, respectively.
The client counts the number $N_{\rm fail}(\le 2k)$ of registers where the client's outcome is different from the server's one.
If $N_{\rm fail}\le 2k\Delta$, where a value of $\Delta$ can be decided by the client, the random-sampling test succeeds.
Finally, the client selects a single register from the remaining $2k$ registers.
The quantum state $\rho_{\rm tgt}$ of the selected register satisfies
\begin{eqnarray}
\langle\Phi^+|\rho_{\rm tgt}|\Phi^+\rangle\ge1-\epsilon+3\Delta-\cfrac{3N_{\rm fail}}{2k}
\end{eqnarray}
with probability at least $1-\delta$~\cite{takeuchi2019quantum}.
Therefore, if the random-sampling test succeeds, i.e., $N_{\rm fail}\le 2k\Delta$, we obtain $\rho_{\rm tgt}$ satisfying Eq.~(\ref{fide2}).

\section{\label{app:c}The uncertainty of the estimation}
The goal of this appendix is to derive the expression of the uncertainty $\delta\omega_{\mathrm{C}}$ ($\delta\omega_{\mathrm{E}}$) in Eq.~\eqref{delomec} (Eq.~\eqref{delomee}).
In our protocol, the standard quantum metrology explained as Appendix \ref{app:a} is implemented with the state preparation error and under the effect of dephasing.
So, in this case, the initial state $\ket{+}$ in the step 1 of Appendix \ref{app:a} is replaced by $\rho(0)=\frac{I}{2}+\sum_{i=x,y,z}\frac{r_i}{2}\sigma_i$ with the fidelity in Eq.~\eqref{fide}.
Also, the evolution of the state in the step 2 of Appendix \ref{app:a} is described by the Hamiltonian $H(t)$ including the effect of dephasing in Eq.~\eqref{ham}.

Let us start by explaining how to calculate the uncertainty if we know the precise form of the initial state.
Based on the average value $S_M$ and the probability $P$ of the outcome $m=1$ for the $\sigma_y$ measurement in Eq.~\eqref{proby}, the qubit frequency $\omega$ can be estimated as follows,
\begin{align}
\omega^{\mathrm{(est)}}_M=\frac{S_M-x(t)}{t y(t)},\label{esto}
\end{align}
where $\omega^{\mathrm{(est)}}_M$ denotes the estimation of the qubit frequency $\omega$ from the average value $S_M$.
The uncertainty $\delta\omega$ of the estimation is defined as the root mean squared error,
\begin{align}
\delta\omega&\equiv \sqrt{\braket{(\omega^{\mathrm{(est)}}_M-\omega)^2}}\label{defun}\\
&=\frac{1}{t y(t)}\sqrt{\braket{(S_M-P)^2}},
\end{align}
where we substitute the estimation $\omega^{\mathrm{(est)}}_M$ of Eq.~\eqref{esto} and the qubit frequency $\omega$ of Eq.~\eqref{proby} for the uncertainty $\delta\omega$.
Since the relation between the variance $\delta^2P$ of the random variable $m$ and the variance of the average value $S_M$ is given as follows,
\begin{align}
\delta^2P=M\braket{(S_M-P)^2},\label{vari}
\end{align}
where $\delta^2P=P(1-P)$, we obtain the uncertainty $\delta\omega$ of the estimation,
\begin{align}
\delta\omega&=\frac{1}{t y(t)}\sqrt{\frac{P(1-P)}{M}}\\
&\simeq \frac{1}{t y(t)}\sqrt{\frac{x(t)\left(1-x(t)\right)}{M}},\label{uncerexa}
\end{align}
where we neglect the term of $\omega t$ due to the assumption of $\omega t\ll1$.

Next, let us explain how to derive the uncertainty $\delta\omega_{\mathrm{C}}$ for the client side.
The client does not know the precise form of the initial state.
However, the client still can assume that the initial state is very close to the ideal state $\ket{+}$ due to the high fidelity guaranteed by the random-sampling test.
So we consider a case that the client tries to estimate the qubit frequency $\omega$ based on the assumption that the initial state is $\ket{+}$.
Of course, in this case, due to the slight deviation of the initial state from $\ket{+}$, there will be systematic errors that cannot be removed just by increasing the repetition number $M$.
By setting $r_x\to 1$ and $r_y\to0$ for Eq.~\eqref{proby}, the probability $P_{\mathrm{C}}$ in the client side for the $\sigma_y$ measurement is given as follows,
\begin{align}
P_{\mathrm{C}}&=P\big|_{\substack{r_x\rightarrow1,\ r_y\rightarrow 0}}\label{probyc}\\
&=x_{\mathrm{C}}+y_{\mathrm{C}}(t)\omega t,
\end{align}
where
\begin{align}
x_{\mathrm{C}}=\frac{1}{2},\quad
y_{\mathrm{C}}(t)=\frac{e^{-g(t)}}{2}.
\end{align}
By using the average value $S_M$ and the probability $P_{\mathrm{C}}$, the qubit frequency is estimated as
\begin{align}
\omega^{\mathrm{(est)}}_{\mathrm{C,}M}=\frac{S_M-x_{\mathrm{C}}}{t y_{\mathrm{C}}(t)}.
\end{align}
Similarly to the above, the uncertainty of the client is defined as the root mean squared error,
\begin{align}
\delta\omega_{\mathrm{C}}&\equiv\sqrt{\braket{(\omega^{\mathrm{(est)}}_{\mathrm{C,}M}-\omega)^2}}\\
&=\frac{1}{t}\sqrt{\frac{\braket{(S_M-P)^2}}{y_{\mathrm{C}}(t)^2}+\left(\frac{P-x_{\mathrm{C}}}{y_{\mathrm{C}}(t)}-\frac{P-x(t)}{y(t)}\right)^2}\\
&\simeq \frac{1}{ty_{\mathrm{C}}(t)}\sqrt{\frac{x(t)\left(1-x(t)\right)}{M}+\left(x(t)-x_{\mathrm{C}}\right)^2}\\
&=\frac{1}{ty_{\mathrm{C}}(t)}\sqrt{\frac{x_{\mathrm{C}}^2}{M}+\left(1-\frac{1}{M}\right)\left(x(t)-x_{\mathrm{C}}\right)^2},\label{delomec}
\end{align}
where we substitute $\omega$ described by the Eq.~\eqref{proby} in the first line, and use $\braket{S_M}=P$ in the second line.
Also, by using Eq.~\eqref{vari} and dropping the term $\omega t$ in the third line, the uncertainty $\delta\omega_{\mathrm{C}}$ of the client is obtained as Eq.~\eqref{delomec}, which is used in the calculation of Eq.~\eqref{uncerc}.

Here, we derive the expression of the uncertainty $\delta\omega_{\mathrm{E}}$ of Eve.
We assume that Eve knows the precise initial state $\rho_{\mathrm{E}}(0)=\frac{I}{2}+\sum_{i=x,y,z}\frac{R_i}{2}\sigma_i$ constrained by the fidelity of Eq.~\eqref{fidee} and Eve can remove the effect of the dephasing.
The probability $P_{\mathrm{E}}$ of Eve for the $\sigma_y$ measurement is calculated as follows,
\begin{align}
P_{\mathrm{E}}&=\mathrm{Tr}\left[e^{-i H_0 t/\hbar}\rho_{\mathrm{E}}(0)e^{i H_0 t/\hbar}\frac{1+\sigma_y}{2}\right]\\
&\simeq x_{\mathrm{E}}+y_{\mathrm{E}}\omega t,\label{probe}
\end{align}
where we use $\omega t\ll 1$ in the second line.
$x_{\mathrm{E}}$ and $y_{\mathrm{E}}$ are defined as,
\begin{align}
x_{\mathrm{E}}=\frac{1+R_y}{2},\quad y_{\mathrm{E}}=\frac{R_x}{2}.
\end{align}
By using the probability $P_{\mathrm{E}}$ and the average value $S_M$, the qubit frequency $\omega$ of Eve can be estimated as,
\begin{align}
\omega_{\mathrm{E,}M}^{\mathrm{(est)}}=\frac{S_M-x_{\mathrm{E}}}{t y_{\mathrm{E}}}.
\end{align}
By repeating the calculation from Eq.~\eqref{defun} to Eq.~\eqref{uncerexa}, we obtain the uncertainty $\delta\omega_{\mathrm{E}}$ of Eve as follows,
\begin{align}
\delta\omega_{\mathrm{E}}&\equiv \sqrt{\braket{(\omega^{\mathrm{(est)}}_{\mathrm{E,}M}-\omega )^2}}\\
&\simeq \frac{1}{t y_{\mathrm{E}}}\sqrt{\frac{x_{\mathrm{E}}\left(1-x_{\mathrm{E}}\right)}{M}}\label{delomee},
\end{align}
where we use the probability $P_{\mathrm{E}}$ in Eq.~\eqref{probe} for $\omega$.
The uncertainty $\delta\omega_{\mathrm{E}}$ of Eve in Eq.~\eqref{uncere} is evaluated based on the expression of Eq.~\eqref{delomee}.

\section{\label{app:d}Contour plots of the uncertainty}
In this Appendix, we show the contour plots of the uncertainty $\delta\omega_{\mathrm{C}}^{(U)}$ in terms of the interaction time $t$ and the repetition number $M$ (or $N$) to investigate the effect of the deviation from the optimized time $t_{\mathrm{opt}}$ on the uncertainty.

\subsection{\label{subsec:slowc}Slow initialization and readout}
\begin{figure*}
  \begin{minipage}[b]{0.45\linewidth}
    \centering
    \includegraphics[width=7.5cm]{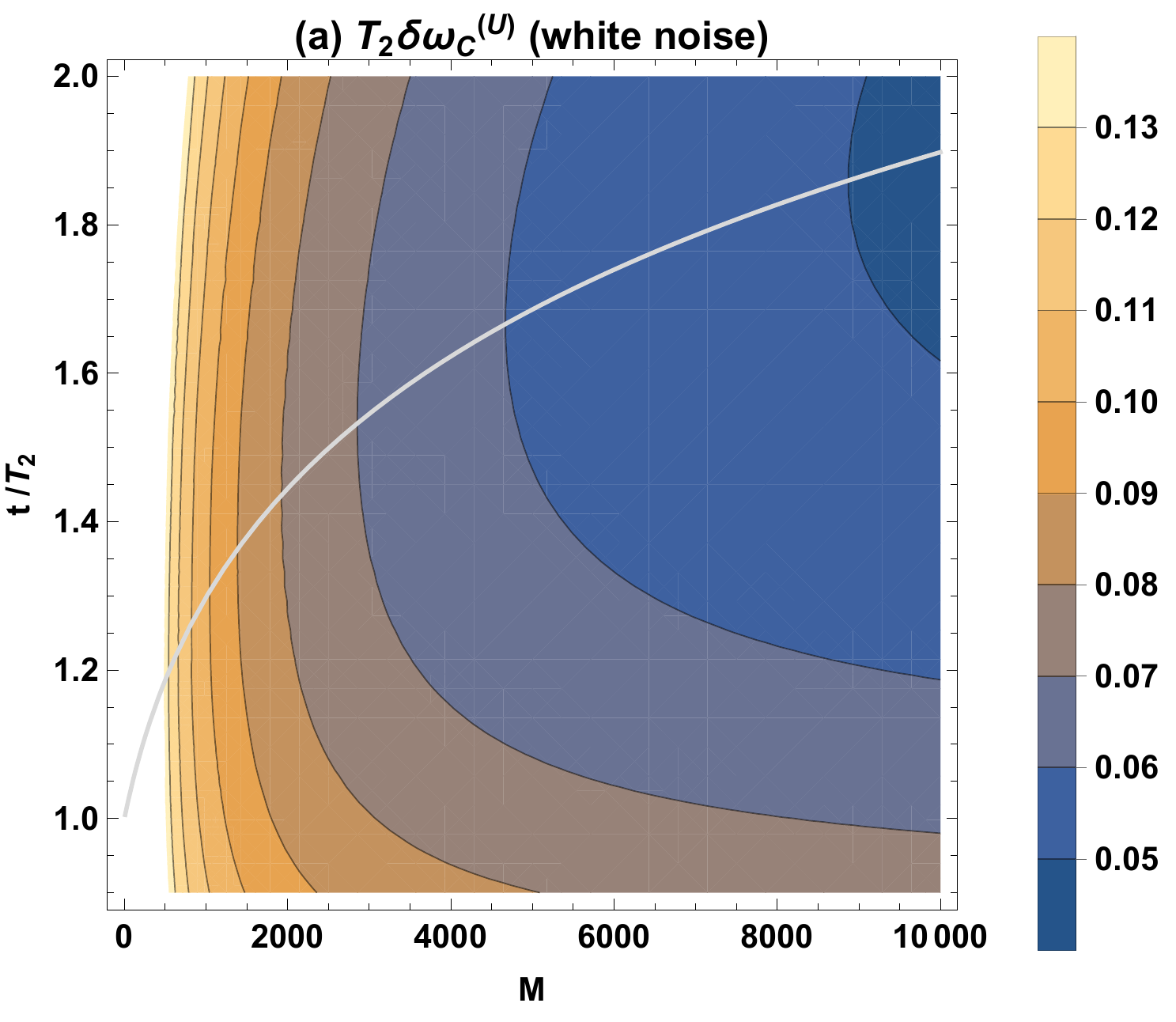}
  \end{minipage}
  \begin{minipage}[b]{0.45\linewidth}
    \centering
    \includegraphics[width=7.5cm]{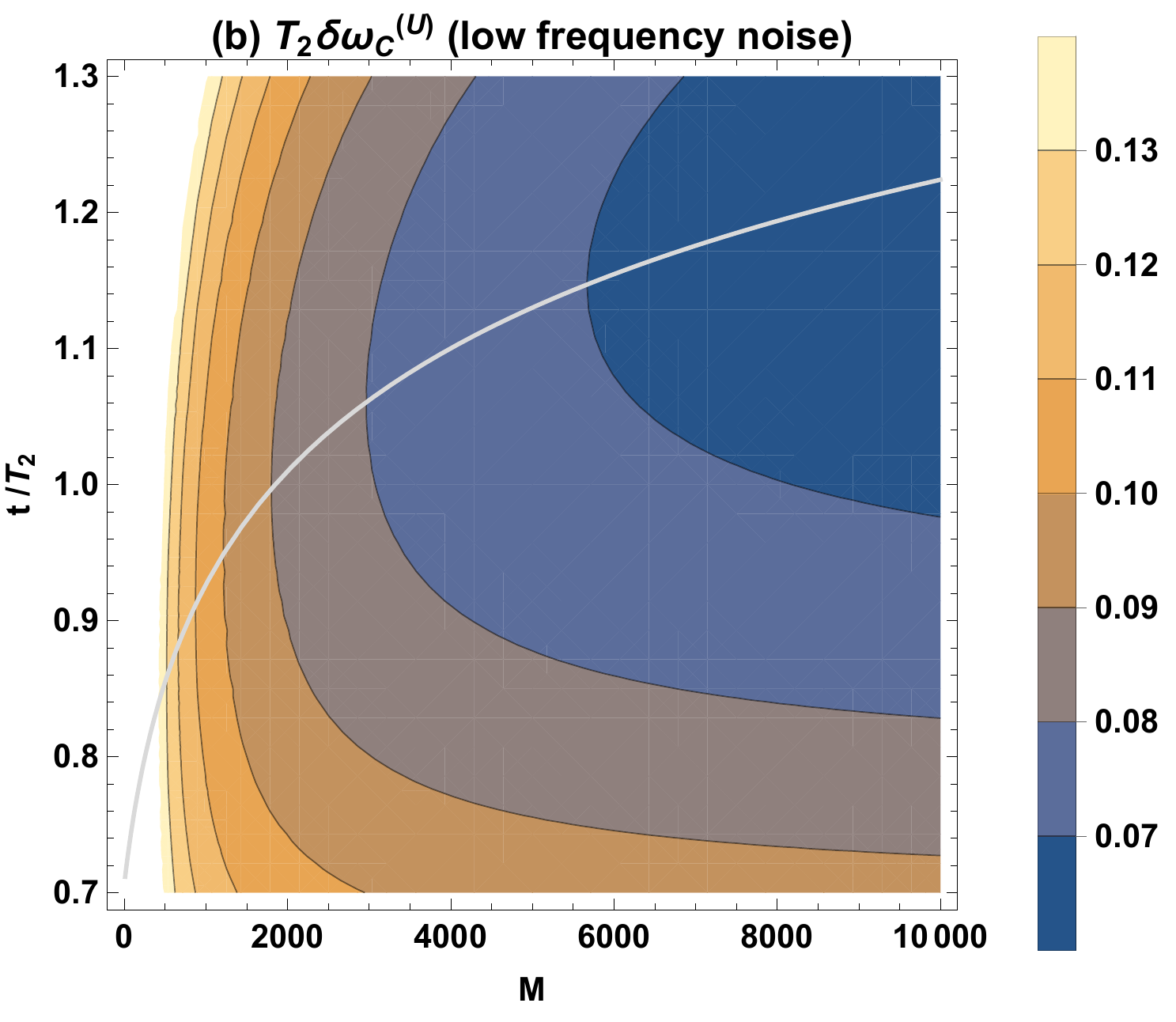}
  \end{minipage}
  \caption{Contour plots of the uncertainty $\delta\omega_{\mathrm{C}}^{(U)}$ of Eq.~\eqref{uncercm} with $\epsilon=0.001$ for (a) the white noise in the range of $M\in(1,10000)$ and $t/T_2\in(0.9,2.0)$, and for (b) low frequency noise in the range of $M\in(1,10000)$ and $t/T_2\in(0.7,1.3)$.
The gray lines denote the optimized time $t_{\mathrm{opt}}$ of Eq.~\eqref{texact}, which are plotted as the solid lines in Fig.~\ref{tplot}.}\label{tmplot}
\end{figure*}
By using Eq.~\eqref{uncercm}, in Fig.~\ref{tmplot}, we show the contour plot of the uncertainty $\delta\omega_{\mathrm{C}}^{(U)}$ with $\epsilon=0.001$ for (a) the white noise in the range of $M\in(1,10000)$ and $t/T_2\in(0.9,2.0)$, and for (b) the low frequency noise in the range of $M\in(1,10000)$ and $t/T_2\in(0.7,1.3)$.
The Fig.~\ref{tmplot} shows that, as we increase the repetition number $M$, the optimized time $t_{\mathrm{opt}}$ gradually increases for both white noise and low frequency noise.
Small fluctuations of the interaction time don't change the uncertainty $\delta\omega_{\mathrm{C}}^{(U)}$ significantly.
This means that a precise timing control is not important for our protocol.

\subsection{\label{subsec:fastc}Fast initialization and readout}
\begin{figure*}
  \begin{minipage}[b]{0.45\linewidth}
    \centering
    \includegraphics[width=7.5cm]{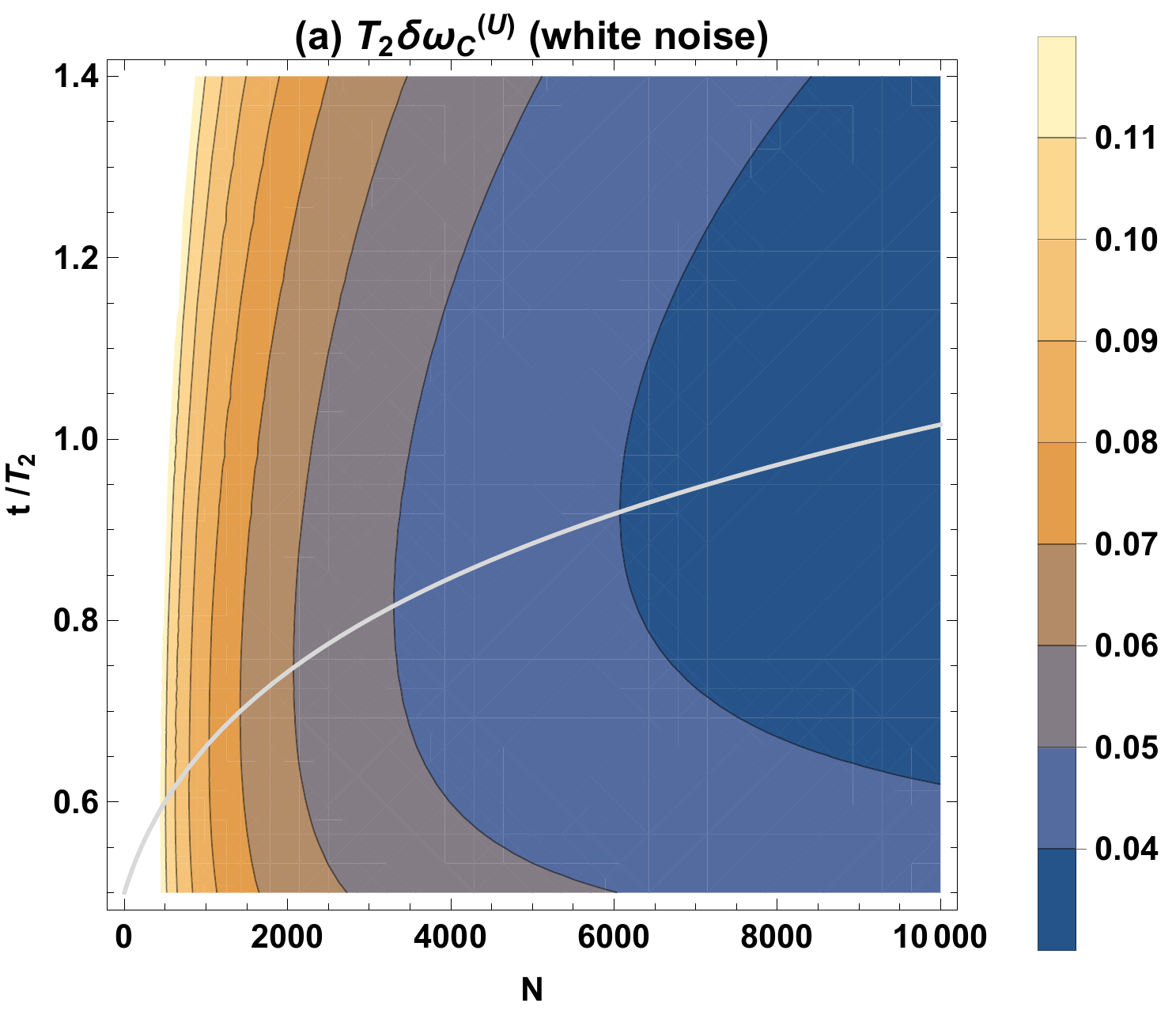}
  \end{minipage}
  \begin{minipage}[b]{0.45\linewidth}
    \centering
    \includegraphics[width=7.5cm]{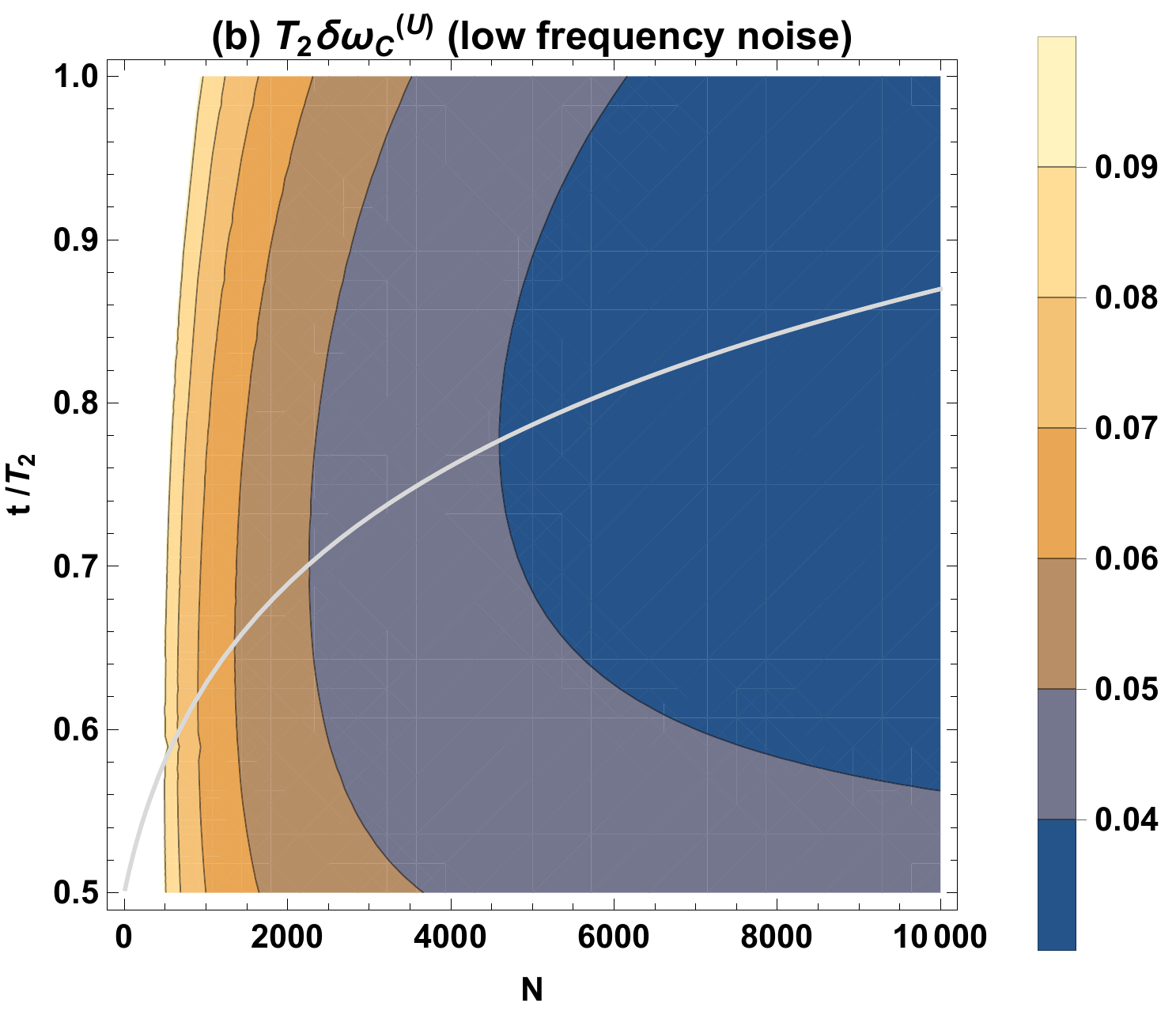}
  \end{minipage}
  \caption{Contour plots of the uncertainty $\delta\omega_{\mathrm{C}}^{(U)}$ of Eq.~\eqref{uncercut} with $\epsilon=0.0001$ for (a) the white noise in the range of $N\in(1,10000)$ and $t/T_2\in(0.5,2.0)$, and for (b) low frequency noise in the range of $N\in(1,10000)$ and $t/T_2\in(0.6,1.4)$.
The gray line denotes the optimized time $t_{\mathrm{opt}}$ to minimize the uncertainty $\delta\omega_{\mathrm{C}}^{(U)}$ for (a) the white noise in Eq.~\eqref{toptuw} and for (b) the low frequency noise in Eq.~\eqref{toptul}, which are plotted as the solid lines in Fig.~\ref{totorplot}.}\label{tnuplot}
\end{figure*}
In Fig.~\ref{tnuplot}, by the use of Eq.~\eqref{uncercut}, we show the contour plots of the uncertainty for (a) the white noise in the range of $N\in(1,10000)$ and $t/T_2\in(0.5,2.0)$ and for (b) the low frequency noise in the range of $N\in(1,10000)$ and $t/T_2\in(0.6,1.4)$ with $\epsilon=0.0001$.
The optimized time $t_{\text{opt}}$ increases against $N$ both in the white noise and the low frequency noise.
Similarly to subsection~(\ref{subsec:slowc}), $\delta\omega_{\mathrm{C}}^{(U)}$ becomes less sensitive to the small change of the interaction time $t$.

Also, since the interaction time $t$ can be optimized for the ratio $\delta\omega_{\mathrm{C}}^{(U)}/\delta\omega_{\mathrm{E}}^{(L)}$ in the case of the fast initialization and readout of Eq.~\eqref{mtind}, we discuss the dependence of the ratio on $N$ and the interaction time $t$.
\begin{figure*}
  \begin{minipage}[b]{0.45\linewidth}
    \centering
    \includegraphics[width=7.5cm]{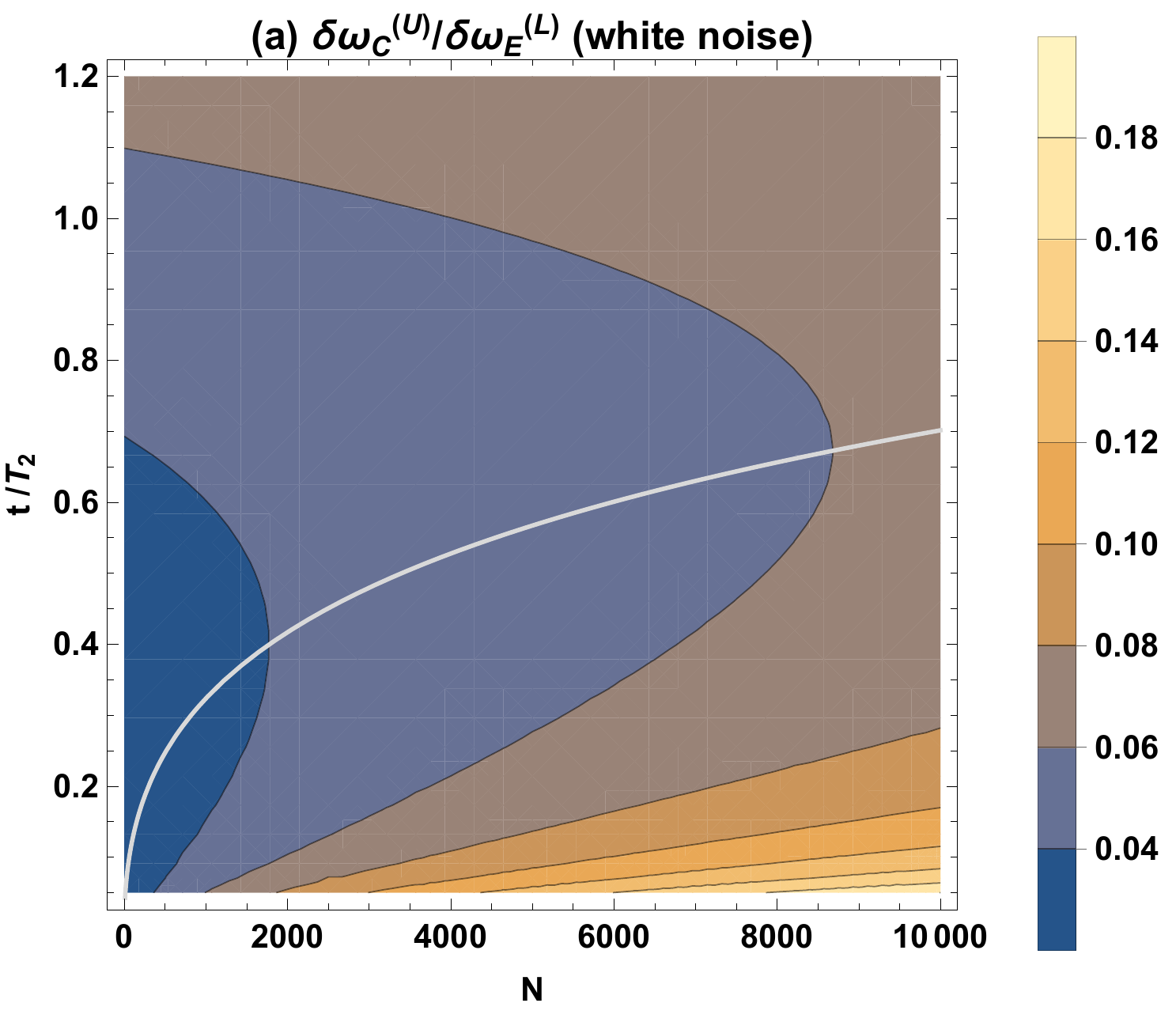}
  \end{minipage}
  \begin{minipage}[b]{0.45\linewidth}
    \centering
    \includegraphics[width=7.5cm]{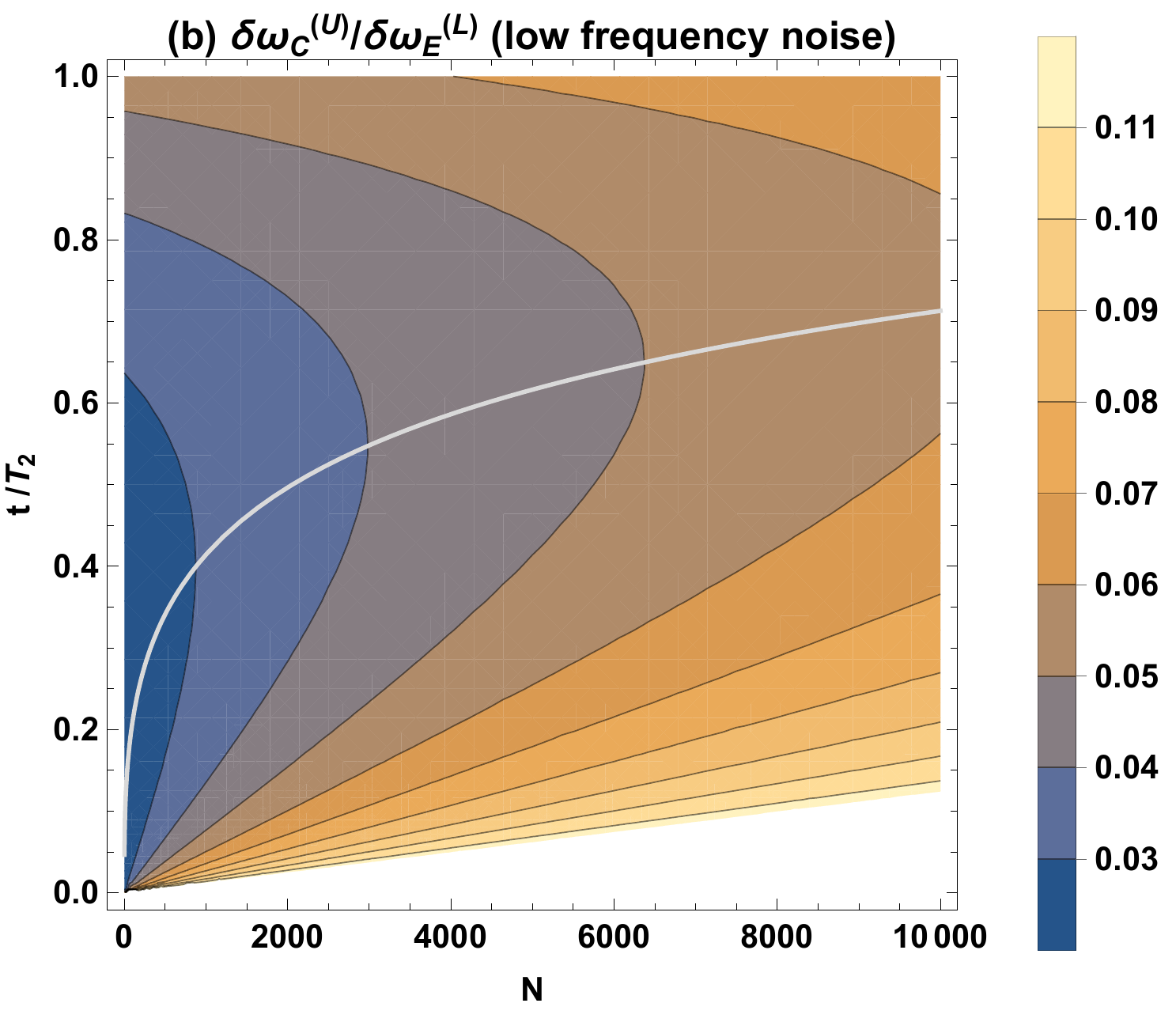}
  \end{minipage}
  \caption{Contour plots of the ratio $\delta\omega_{\mathrm{C}}^{(U)}/\delta\omega_{\mathrm{E}}^{(L)}$ of Eq.~\eqref{ratiot} with $\epsilon=0.0001$ for (a) the white noise in the range of $N\in(1,10000)$ and $t/T_2\in(0.2,2.0)$, and for (b) low frequency noise in the range of $N\in(1,10000)$ and $t/T_2\in(0.2,1.4)$.
The gray line denotes the optimized time $t_{\mathrm{optR}}$ of Eq.~\eqref{toptr} to minimize the ratio $\delta\omega_{\mathrm{C}}^{(U)}/\delta\omega_{\mathrm{E}}^{(L)}$ for (a) the white noise and for (b) the low frequency noise, which are plotted as the dashed lines in Fig.~\ref{totorplot}.}\label{tnrplot}
\end{figure*}
In the left side of Fig.~\ref{tnrplot}, we show the contour plot of the ratio $\delta\omega_{\mathrm{C}}^{(U)}/\delta\omega_{\mathrm{E}}^{(L)}$ of Eq.~\eqref{ratiot} for the white noise with $\epsilon=0.0001$ in the range of $N\in (1,10000)$ and $t/T_2\in (0.2, 2.0)$.
Similarly, in the right side of Fig.~\ref{tnrplot}, we numerically calculate the ratio $\delta\omega_{\mathrm{C}}^{(U)}/\delta\omega_{\mathrm{E}}^{(L)}$ of Eq.~\eqref{ratiot} for the low frequency noise with $\epsilon=0.0001$ in the range of $N\in (1,10000)$ and $t/T_2\in (0.2, 1.4)$.
The ratio $\delta\omega_{\mathrm{C}}^{(U)}/\delta\omega_{\mathrm{E}}^{(L)}$ similar to the uncertainty $\delta\omega_{\mathrm{C}}^{(U)}$ doesn't change significantly for the small deviation from the optimized time $t_{\mathrm{optR}}$.


\bibliography{ref} 
\bibliographystyle{apsrev4-2}

\end{document}